\documentclass[%
 aip,
 amsmath,amssymb,
 reprint,%
]{revtex4-2}

\usepackage{graphicx}% %Include figure files
\usepackage{dcolumn}% Align table columns on decimal point
\usepackage{bm}% bold math
\usepackage[utf8]{inputenc}
\usepackage[colorlinks]{hyperref}
\usepackage{float}
\usepackage{amsmath}
\usepackage{xspace}
\usepackage{amsmath}
\usepackage{mathrsfs}
\usepackage{xcolor}

\newcommand{\ep}{\epsilon}

\newcommand\Tstrut{\rule{0pt}{2.3ex}}
\newcommand{\NbN}{NbSi$_2$N$_4$\xspace}

\newcommand{\VN}{VSi$_2$N$_4$\xspace}
\newcommand{\VP}{VSi$_2$P$_4$\xspace}
\newcommand{\VGN}{VGe$_2$N$_4$\xspace}
\newcommand{\VGP}{VGe$_2$P$_4$\xspace}
\newcommand{\MoN}{MoSi$_2$N$_4$\xspace}

\newcommand{\TaGN}{TaGe$_2$N$_4$\xspace}
\newcommand{\dzs}{d$_{z^2}$\xspace}
\newcommand{\dxsys}{d$_{x^2-y^2}$\xspace}

\begin{document}

\title[]{Magnetic Properties of \NbN, \VN, and \VP Monolayers}% Force line breaks with \\

\author{Md. Rakibul Karim Akanda}
\thanks{makan001@ucr.edu}
\affiliation{Laboratory for Terahertz and Terascale Electronics (LATTE), Department of Electrical and Computer Engineering, University of California, Riverside, CA 92521, USA}

\author{Roger K. Lake}
\thanks{Corresponding author: rlake@ece.ucr.edu}
\affiliation{Laboratory for Terahertz and Terascale Electronics (LATTE), Department of Electrical and Computer Engineering, University of California, Riverside, CA 92521, USA}

\begin{abstract}

The recent demonstration of \MoN and its exceptional stability to air, water, acid, and heat
has generated intense interest
in this family of two-dimensional (2D) materials.
Among these materials, monolayers of \NbN, \VN, and \VP are semiconducting, easy-plane ferromagnets
with negligible in-plane magnetic anisotropy.
They thus satisfy a necessary condition for exhibiting a dissipationless spin superfluid mode.
The Curie temperatures of monolayer \VP and \VN are determined to be above room temperature
based on Monte Carlo and density functional theory calculations.
The magnetic moments of \VN can be switched from in-plane to out-of-plane by applying
tensile biaxial strain or electron doping.
\end{abstract}

\maketitle

Two dimensional (2D) layered materials with transition metals have been of interest for many decades due to the
correlated phenomena and multiple polymorphs and phases that they can
exhibit such as charge density waves,\cite{Yoffe87}
superconductivity,\cite{Klemm_Layered_SCs} and magnetism.\cite{Jongh_Mag_Props_Layered_Mats}
The ability to exfoliate or grow single monolayers
renewed the interest in these materials for possible electronic and
optoelectronic applications\cite{2011_ML_MoS2_FET_AKis,2020_2D_Review_2DMat}
by both traditional transistor type devices\cite{wang2012electronics,Fiori_2DMat_Review_NNano14}
and by exploiting external control of their phase transitions.\cite{2021_Phase_Transitions_NatRev}
The relatively recent demonstration of magnetism in single monolayer of CrI$_3$\cite{CrI3}
and bilayers of Cr$_2$Ge$_2$Te$_6$\cite{2017_CGT_Nat}
has spurred intense experimental and theoretical activity to find other 2D magnetic materials with higher transition
temperatures.\cite{2018_Fe3GeTe2_FM_Gate_Tunable_Nat,
2019_2D_Mag_Novoselov_Rev_NMat,
2019_2D_Mag_Rev_AdvMat,
2020_MAX3_FM_MacDonald_PRB,
2020_Hi_Thruput_2D_FM_NPJCM,
2020_CrSeX_FM_JPCehmC,
2020_FGT_PSS_Fokwa,
2020_2D_Mag_Mats_NanoToday}

The most recent addition to the family of 2D materials are the transition metal silicon nitrides, phosphides,
and arsenides with the chemical formulas MA$_2$Z$_4$, where M is the transition metal,
A $\in$ \{Si, Ge\}, and Z $\in$ \{N, P, As\}.\cite{2020_MSi2N4_Sci}
High quality multilayers and monolayers
of \MoN were grown using chemical vapor deposition, and what was particularly notable
was their stability to air, water, acid, and heat that was unprecedented among transition
metal 2D materials.\cite{2020_MSi2N4_Sci}
This rather mundane property is highly desirable for manufacturing applications.
While BN encapsulation is an effective solution for stabilizing reactive 2D materials for
laboratory experiments,\cite{2014_BP_JLau_2DMats} it is less than ideal for manufacturing.
Only \MoN was experimentally characterized in detail, WSi$_2$N$_4$ was also grown, and
12 materials were simulated with density functional theory (DFT) and found to be stable.
Among these 12, two of the nitrides, \VN and \NbN, were identified as magnetic.

\begin{figure}
\includegraphics[width=0.45\textwidth]{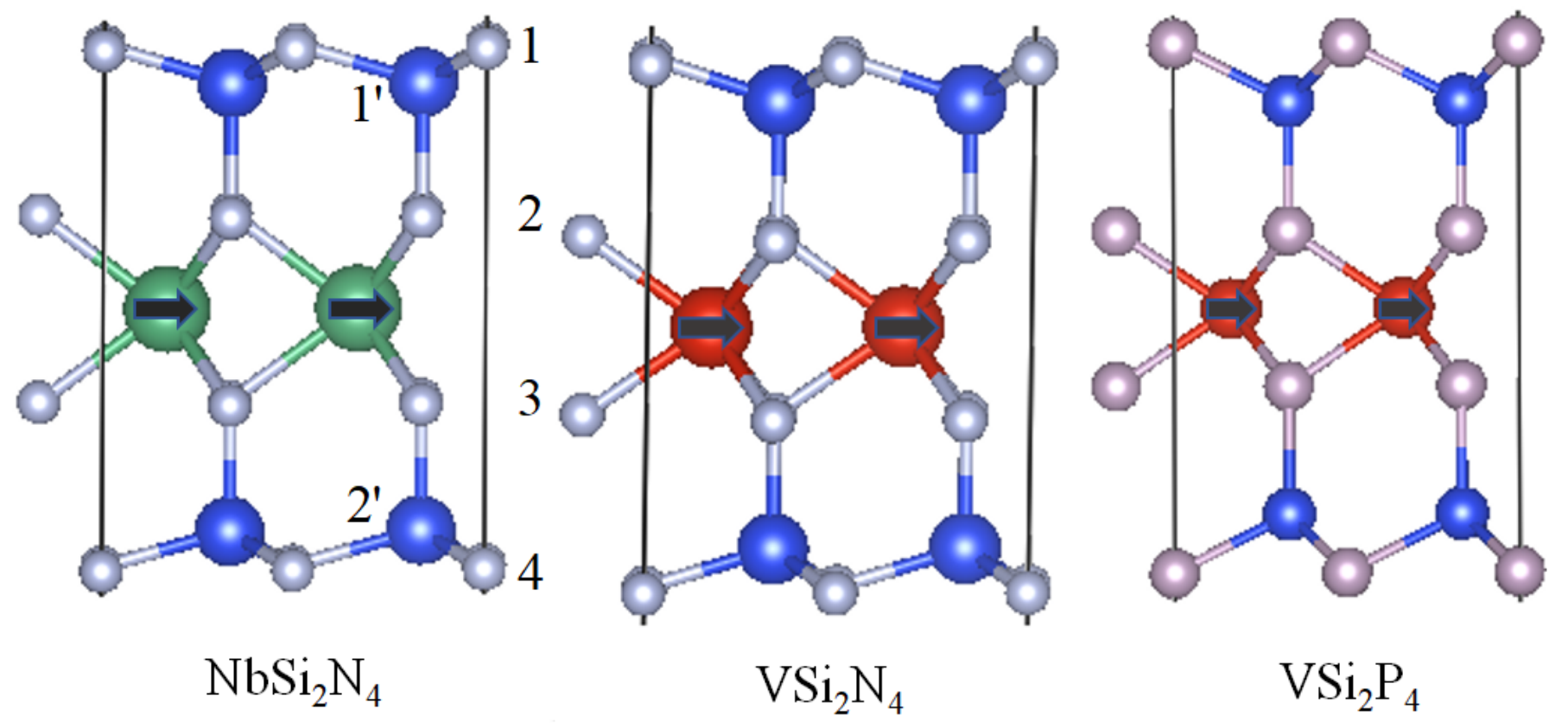}
\caption{Structures and magnetic orientations of $\alpha_1$-\NbN, $\alpha_1$-\VN, and $\alpha_1$-\VP.
Blue atoms are Si, white atoms are N, and pink atoms are P.
The directions of the magnetic moments are shown on the transition metals Nb and V.
All materials are easy-plane, semiconducting ferromagnets.
}
\label{fig:Structures}
\end{figure}

This work motivated immediate follow-on theoretical investigations of this material family
both determining properties of the materials and extending the list of stable materials.
\cite{
2020_VSi2P4_FM_Piezo_PCCP,
2020_MoSi2N4_Contacts_APL,
2020_Valley_Dep_Properties_PRB,
2021_MA2Z4_TI_to_Ising_SC_NComm,
2020_MoSi2N4_ML_properties_arxiv,
2021_MoSi2N4_Kappa_NJP,
2021_MoSi2N4_WSi2N4_MIT_APL,
2021_MA2Z4_MIT_Strain_PRB,
2021_MoSe2Z4_opto_Nanomat}
The most extensive theoretical survey found 32 thermodynamically and dynamically stable compounds of the form MA$_2$Z$_4$.
\cite{2021_MA2Z4_TI_to_Ising_SC_NComm}
Of these, 6 were magnetic: \VN, \VP, \NbN, \VGN, \VGP, and \TaGN.
Based on the formation enthalpies, a Si based compound MSi$_2$Z$_4$ is approximately 3 times more stable
than its equivalent Ge based compound MGe$_2$Z$_4$,
and a silicon nitride compound MSi$_2$N$_4$ is also approximately 3 times more stable than its equivalent
silicon phosphide compound MSi$_2$P$_4$.

Other recent works investigated
MSi$_2$Z$_4$ bilayers (M = Ti, Cr, Mo; Z = N, P)
for their sensitivity to vertical strain
\cite{2021_MA2Z4_MIT_Strain_PRB,2021_MoSi2N4_WSi2N4_MIT_APL},
and MoSi$_2$N$_4$ and WSi$_2$N$_4$
for their sensitivity to biaxial strain and
vertical applied electric field.\cite{2021_MoSi2N4_WSi2N4_MIT_APL}
It was found that vertical strain can cause an insulator to metal transition;
\cite{2021_MA2Z4_MIT_Strain_PRB}
biaxial strain can lead to an indirect to direct bandgap transition;
\cite{2021_MoSi2N4_WSi2N4_MIT_APL}
and that an applied electric field can result in an insulator to metal transition.
\cite{2021_MoSi2N4_WSi2N4_MIT_APL}
A theoretical investigation of 2D-2D contacts to monolayer MoSi$_2$N$_4$ using NbS$_2$ and
graphene found ultralow p-type Schottky barriers with NbS$_2$ contacts and approximately
equal n-type and p-type Schottky barriers with graphene contacts.
\cite{2020_MoSi2N4_Contacts_APL}
The graphene Schottly barriers were shown to be tunable with an applied vertical electric field.

In this letter, we theoretically
investigate the magnetic properties of the three Si based magnetic materials: \VN, \VP, \NbN.
Values of the exchange constants and magnetic anisotropy energies are determined,
and the Curie temperatures are calculated.
The Curie temperatures of \VN and \VP are near or at room temperature and above,
depending on the model used, whereas the Curie temperature of \NbN is low.
Therefore, the primarily focus will be on the two vanadium compounds.
The magnetic anisotropy energies of \VN and \VP
are calculated as a function of uniaxial and biaxial strain and electron and hole doping.

These materials can exist in a variety of hexagonal phases.
Ref. [\onlinecite{2021_MA2Z4_TI_to_Ising_SC_NComm}] calculated the formation energies of 30 different phases
and found that the lowest energy phase of \NbN and \VN was the $\alpha_1$ phase, which is the same phase
as the lowest energy phase of \MoN.
The formation energy of the next higher energy phase ($\delta_4$) of \NbN
was 13 meV per atom higher than the $\alpha_1$ phase,
and the formation energy of the next higher energy phase ($\beta_2$) of \VN
was 6 meV per atom higher than the $\alpha_1$ phase.
For \VP, the lowest energy phase was $\delta_4$ with a formation energy 0.3 meV
below that of $\alpha_1$.
Since the formation energy of the $\alpha_1$ phase of \VP is so close to that of the $\delta_4$ phase,
and since the $\delta_4$ phase has been previously analyzed \cite{2021_MA2Z4_TI_to_Ising_SC_NComm},
we will consider the $\alpha_1$ phase of all 3 materials: \NbN, \VN, and \VP.
We note that the piezoelectric and magnetic properties of the $\alpha_1$ phase of \VP have recently been
investigated with DFT using
the generalized gradient approximation (GGA).\cite{2020_VSi2P4_FM_Piezo_PCCP}

%-----------------------------------Method------------------------------------------------------------------

DFT combined with Monte Carlo (MC) calculations are applied
to determine the electronic and magnetic properties of these materials.
The magnetic anisotropy energy (MAE) and exchange energy are evaluated from DFT
calculations implemented in the Vienna ab initio simulation package (VASP).\cite{VASP}
The electron-core interactions are described by the projected augmented wave (PAW) potentials.\cite{PAW}
Electronic structure is calculated using three different functionals:
GGA as parameterized by Perdew-Burke-Ernzerhof (PBE), PBE plus the Hubbard U correction (PBE+U),
and the Heyd-Scuseria-Ernzerhof hybrid functional (HSE06).\cite{PBE,PBEU,HSE03,HSE06}
The PBE+U calculation includes a Hubbard U correction term $U_{\rm eff} = 1$ eV for the V atom
where $U_{\rm eff} = U - J$.\cite{HubbardUAPL,HubbardPRB}
For calculations of the MAE, spin orbit coupling (SOC) must also be included.
All plots shown for the MAE are the results from PBE(SOC)+U calculations.
The cutoff energies for expanding the plane wave basis are 600 eV.
Integration over the Brillouin zone uses a Monkhorst–Pack scheme
with a $\Gamma$-centered $16\times 16 \times 1$ k-point grid, an energy broadening
parameter of 50 meV, and the total energy is converged to $10^{-6}$ eV.
Structures are relaxed until the forces are less than 0.001 eV/\AA.
A vacuum spacing of 20 {\AA} is used in the direction normal to the
2D monolayer to eliminate the interactions from periodic images.
The calculated lattice constants for \NbN, \VN, and \VP are 2.96 {\AA}, 2.88 \AA, and 3.47 \AA,
respectively,
and they are very close to previous reported results.\cite{2020_MSi2N4_Sci,2020_VSi2P4_FM_Piezo_PCCP}

Uniaxial strain is applied along the $x$ axis corresponding to lattice vector $a_1$.
The applied strain is evaluated using $\epsilon= (a-a_0)/a_0 \times 100\%$,
where $a$ and $a_0$ are the lattice parameters
of the strained and unstrained monolayer.
Biaxial strain is applied by uniformly varying both in-plane lattice constants.
For each strain, the atomic positions are relaxed using the DFT parameters described above.
Spin-polarized self consistent calculations are performed with the relaxed structure for each strain to
obtain the charge density.
Using the charge densities, total energies are calculated in the presence of
spin orbit coupling (SOC) for in-plane ($E_{||}$) and out-of-plane ($E_\perp$) magnetization
to find the magnetic anisotropy energy (MAE).
The MAE is defined as $E_{\rm MAE} = E_{\perp} - E_{||}$.
Positive magnetic anisotropy indicates that in-plane magnetization is favored.
The values provided in meV are per magnetic atom, which for these 3 materials under consideration,
are also per unit cell.

To investigate the transition temperatures,
a nearest neighbor Heisenberg type Hamiltonian with magnetic anisotropy and long range dipole-dipole interactions
is constructed,
\begin{equation}
H = -\tfrac{1}{2} J \sum_{\langle i,j \rangle} {\bf S}_i \cdot {\bf S}_j + k_u \sum_i \left( S^z_i \right)^2 + H_{dd},
\end{equation}
where ${\bf S}_i$ is the spin for magnetic atom $i$,
$\langle i,j \rangle$ are the indices of nearest neighbor magnetic atoms,
$k_u > 0$ is the easy-plane magnetic anisotropy energy per magnetic atom,
and $H_{dd}$ is the dipole-dipole interaction.
The exchange energies ($J$) of these materials are calculated from the total energy differences
between the antiferromagnetic ($E_{AFM}$) and ferromagnetic ($E_{FM}$) states.\cite{diffapproach,diffapproachref,j1j2j3,j12}
With six nearest neighbours in the monolayer of 2D hexagonal lattice,
the nearest-neighbor exchange energy is given by,\cite{j1j2j3,j12,j13}
$J=( E_{\rm AFM} - E_{\rm FM} ) / 12$.
In these monolayer materials all of the magnetic atoms are in the same plane.
The second neighbor exchange energies were previously determined for VSi$_2$P$_4$ and
found to be negligible compared to the nearest neighbor energies, \cite{2021_MA2Z4_TI_to_Ising_SC_NComm}
and, therefore, they are ignored.
The Curie temperatures are determined from
Monte Carlo (MC) calculations as implemented in the VAMPIRE software package
using a $10 \times 10$ supercell.\cite{vampire_code,vampire_JPCM14}
The MC calculations incorporate the magnetic anisotropy energies, the magnetic moments,
the exchange energies, lattice constants, and atomic positions determined from the DFT calculations, and they also include the long range dipolar interactions.
This MC approach has been used to determine the Curie temperatures in other monolayer 2D materials such
as \VP, VI$_3$, CrSBr, and CrSeBr.
\cite{2021_MA2Z4_TI_to_Ising_SC_NComm,
VI3_Tc_JPCM20,
CrSBrandCrSeBr}
We note that the Curie temperatures in other monolayer 2D materials determined from the
MC approach have been shown to compare
well to temperatures determined from renormalized spin wave theory (RSWT).
For example, the MC/RSWT predicted Curie temperatures for monolayer
CrSBr and CrSeBr were 168/150 K and 150/152 K, respectively.\cite{CrSBrandCrSeBr}
%

%*************** E-ks *************************
\begin{figure}
\includegraphics[width=1.0\linewidth]{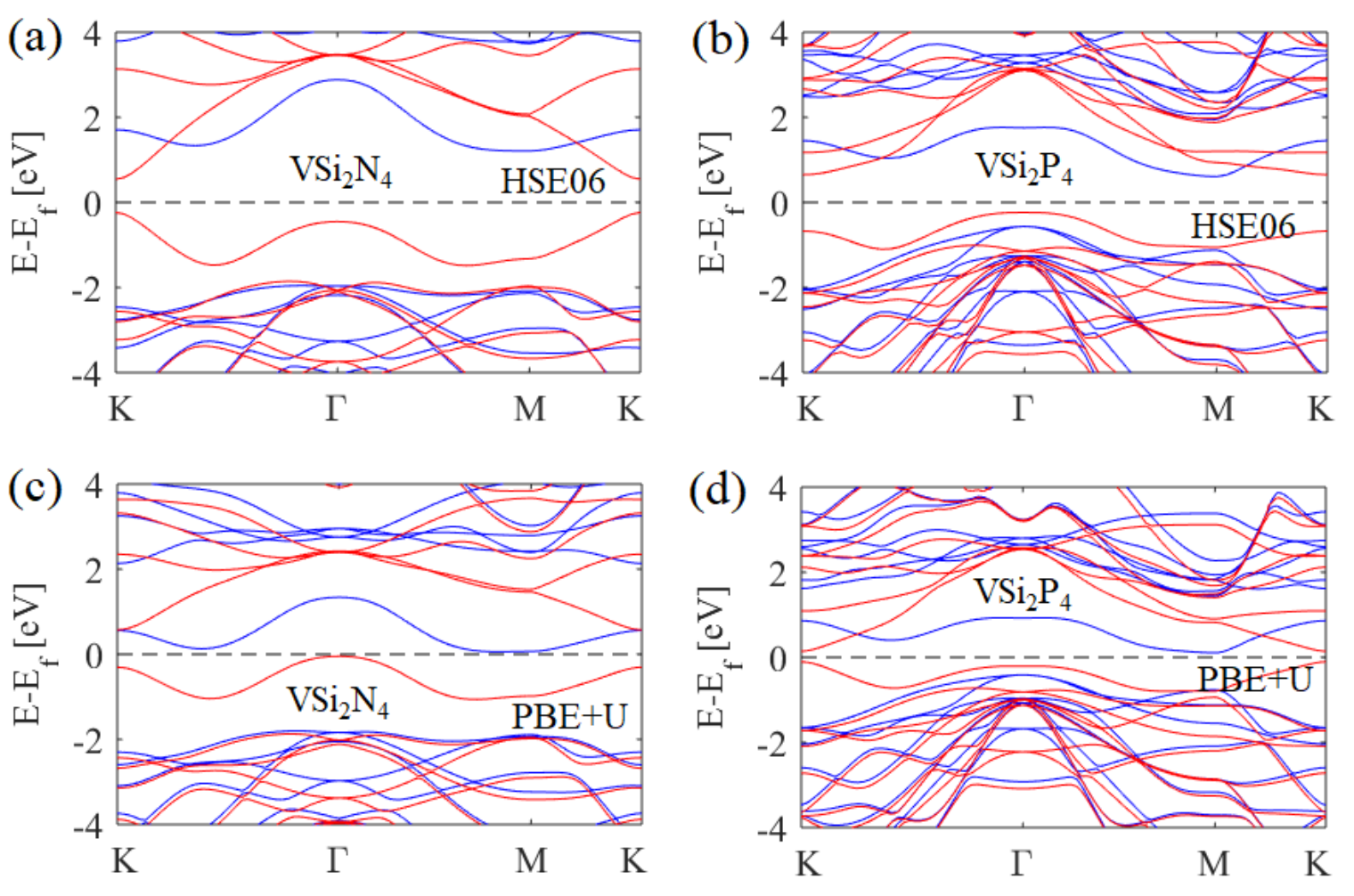}
\caption{Spin resolved energy bands of \VN and \VP calculated with (a,b) HSE06
and (c,d) PBE+U.
Spin up bands are red and spin down bands are blue.
The two spin-split, narrow bands on either side of the Fermi level are d-orbital bands
centered on the transition metal.
}
\label{fig:EkplainUpDown}
\end{figure}

%***********************************************************************************************************************************
%---------------------------------------------------------- Results ----------------------------------------------------------------
%***********************************************************************************************************************************
%

%
The structures and the ground state orientations of the magnetic moments of
\NbN, \VN, and \VP are shown in Fig.~\ref{fig:Structures}, and the spin-resolved band structure
for \VN and \VP,
calculated with the HSE06 and PBE+U functionals, are shown in Fig. \ref{fig:EkplainUpDown}.
The band structure for all three materials calculated with PBE, PBE+U, and HSE06
are shown in the Fig. S3 of the Supplement, and the d-orbital resolved bands of \VN
and \VP are shown in Fig. S2.
For all 3 materials at $\Gamma$,
the two isolated narrow bands near the Fermi level are
spin-polarized, ${\rm d}_{z^2}-{\rm orbital}$ bands centered on the transition metal atoms.
Near K, these bands transition to d$_{x^2-y^2}$.
The next higher conduction band is primarily d$_{x^2-y^2}$ at $\Gamma$, and it transitions to d$_{z^2}$
at K.
The electronic structures from the HSE06 calculations show the following properties.
In \VN and \VP, the higher spin-up conduction band at $\Gamma$ crosses the lower spin-down band near
K where the orbital composition of the two bands switch, \dzs $\Longleftrightarrow$ \dxsys.
\VN is direct gap (0.78 eV) at K with a \dxsys spin-up valence band and a \dzs spin-up conduction band.
\VP is indirect gap (0.84 eV) with the d$_{z^2}$ spin-up valence band edge at $\Gamma$ and a mixed
d$_{z^2}$ + \dxsys spin-down conduction band edge at M (the spin-up band at K is 40 meV higher) .
\NbN is indirect gap (0.54 eV) with the spin-up d$_{z^2}$ valence band at $\Gamma$ and a
mixed orbital \dzs + \dxsys spin-down conduction band near M.
The lower valence bands are primarily p-orbital bands which come from the N and P atoms.
At the PBE level of theory, all 3 materials are semi-metals.
Adding the Hubbard U correction creates a small gap at the Fermi level for \VN and \VP.
Adding a percentage of exact exchange with the HSE06 functional increases the gap substantially.
We note that the HSE06 calculations match those reported in the Supplement of
Ref. [\onlinecite{2021_MA2Z4_TI_to_Ising_SC_NComm}].

The equilibrium \NbN, \VN, and \VP monolayers are easy-plane, semi-conducting ferromagnets.
The calculated equilibrium magnetic moments, exchange energies, magnetic anisotropy energies,
and Curie temperatures are shown in Table \ref{tab:mag_values_all} for different levels of theory.
The magnetic moments
are comparable to those from prior studies.\cite{2020_MSi2N4_Sci,2020_VSi2P4_FM_Piezo_PCCP}
The positive MAE values indicate in-plane alignment of the magnetic moments.
Within the energy resolution of our calculations (1 $\mu$eV),
the total energy is independent of the angle of the magnetic moment within the plane of the monolayer.
Since the monolayers are insulating, easy-plane FMs with extremely weak in-plane anisotropy,
they satisfy the necessary conditions for exhibiting a dissipationless spin superfluid mode.
\cite{2014_SSF_YT_PRL}
%
%
%*************** Table J with PBE, PBE+U and HSE *************************
\begin{table}
\begin{tabular}{p{2.7cm}p{1.5cm}p{1.5cm}p{1cm}p{1.3cm}}
\hline
\Tstrut
Materials (Theory) & Magnetic Moment ($\mu_{B}$) & Exchange energy ($\times 10^{-21}$J) & MAE ($\frac{\rm MJ}{\rm m^3}$) & $T_C$ (K)\\
\hline
\hline
\NbN (PBE) &  $0.32$ & $0.064$ & $0.30$ & $12$\\
\VN (PBE)  & $0.93$ & $1.50$  & $0.24$ & $301$\\
\VP (PBE)  & $0.96$ & $1.11$   & $0.14$ & $235$\\
\VN (PBE+U) & $1.05$ & $2.53$   & $0.25$ & $452$\\
\VP (PBE+U) & $1.04$ & $1.77$  & $0.11$ & $350$\\
\VN (HSE06) & $1.19$ & $2.80$  & $--$  & $^*506$\\
\hline
\end{tabular}
	\caption{Magnetic moment, exchange energy per link, equilibrium magnetic anisotropy energy (MAE), and
Curie temperature, calculated using different functionals
for \NbN, \VN, and \VP.
$^*T_C$ is calculated using the HSE06 exchange energy and the PBE(SOC)+U MAE.}
\label{tab:mag_values_all}
\end{table}

\begin{figure}
\includegraphics[width=\linewidth]{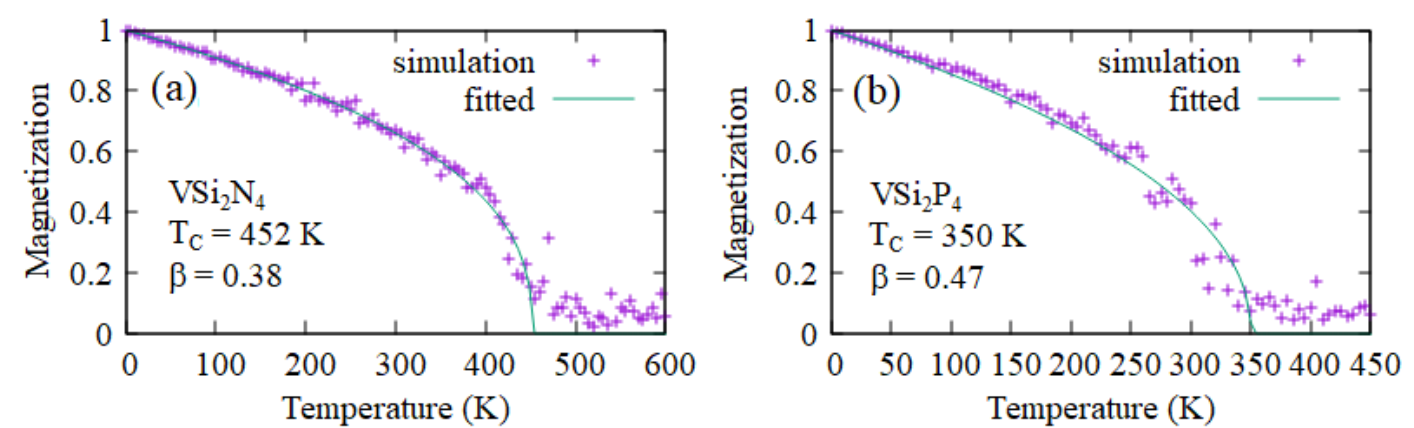}
\caption{Monte Carlo calculations of the normalized magnetization as a function of temperature for
(a) VSi\textsubscript{2}N\textsubscript{4} and
(b) VSi\textsubscript{2}P\textsubscript{4}.
The solid lines show the best fits to the analytical expression given by Eq. (\ref{eq:mT}).
}
\label{fig:Curie}
\end{figure}

The normalized magnetizations, determined from Monte Carlo calculations,
are plotted as a function of temperature for \VN and \VP
in Fig.~\ref{fig:Curie}.
The solid lines show the best fits to the analytic expression,
\begin{equation}
m(T) = (1-T/T_C)^\beta .
\label{eq:mT}
\end{equation}
The fitted values of $T_C$ and $\beta$ are shown on the plots.
At the PBE+U level of theory, both monolayer \VP and \VN have Curie temperatures
above room temperature, 350 K and 452 K, respectively.
The effect of the increasing levels of theory, PBE, PBE+U, and HSE06, is to successively increase the
bandgap between the two spin-polarized d$_{z^2}$ bands, so that at the HSE06 level, the two bands are completely
gapped which maximizes the spin polarization, magnetic moment, and the exchange constant.
The predicted $T_C$ increases with the increasing gap, giving a maximum value of 506 K for \VN
using the exchange constant determined from the HSE06 calculation.
Since \VN has the same structure, surface chemistry, and formation energy\cite{2021_MA2Z4_TI_to_Ising_SC_NComm}
as the experimentally
characterized \MoN, we expect \VN to also be an air/water-stable material with
$T_c > 100^\circ$ C, which is a criterion for operation in a modern integrated circuit environment.
%
%*************** M versus T *************************

For context, we compare to several other 2D materials with similar
predicted Curie or N\'{e}el temperatures.
The calculated Curie-temperatures for monolayers of the transition metal dichalcogenides (TMDs)
VS\textsubscript{2}, VSe\textsubscript{2}, and VTe\textsubscript{2}
are $292$ K, $472$ K, and $553$ K, respectively.\cite{VS2andVSe2andVTe2}
The predicted monolayer Curie temperatures for the ternary 2D materials
CrSeCl, VSeTe, and CrSeI are
$320$ K, $350$ K, and $360$ K, respectively.\cite{VSeTe,CrSBrandCrSeBr,2020_CrSeX_FM_JPCehmC}
Monolayers of ferromagnetic semiconductors TcSiTe\textsubscript{3}
have a predicted Curie temperature of $538$ K.\cite{MGeTe3}
We note that Tc is radioactive, so that Tc compounds are unlikely to see magnetic applications.
Monolayers of
RuI\textsubscript{3},
MnN,
Co\textsubscript{2}S\textsubscript{2},
and
3R-MoN\textsubscript{2}
have Curie temperatures of
$360$ K,
$368$ K,
$404$ K,
and
$420$ K,
respectively.\cite{Co2S2,RuI3,MoN2,MnN}
We note that synthesis of layered 3R-MoN$_2$ requires high pressure,\cite{2015_3R-MoN2_Hi_P_Synthesis_JACS}
which is not amenable to thin film growth techniques, and the proposed graphitization synthesis route to achieve MnN
monolayers\cite{2014_Spontaneous_Graphitization} has not yet been demonstrated.
Within the list above, \VN stands out for its combination of air stability and relatively high Curie temperature.
%*************** MAE versus strain *************************
\begin{figure}
\includegraphics[width=\linewidth]{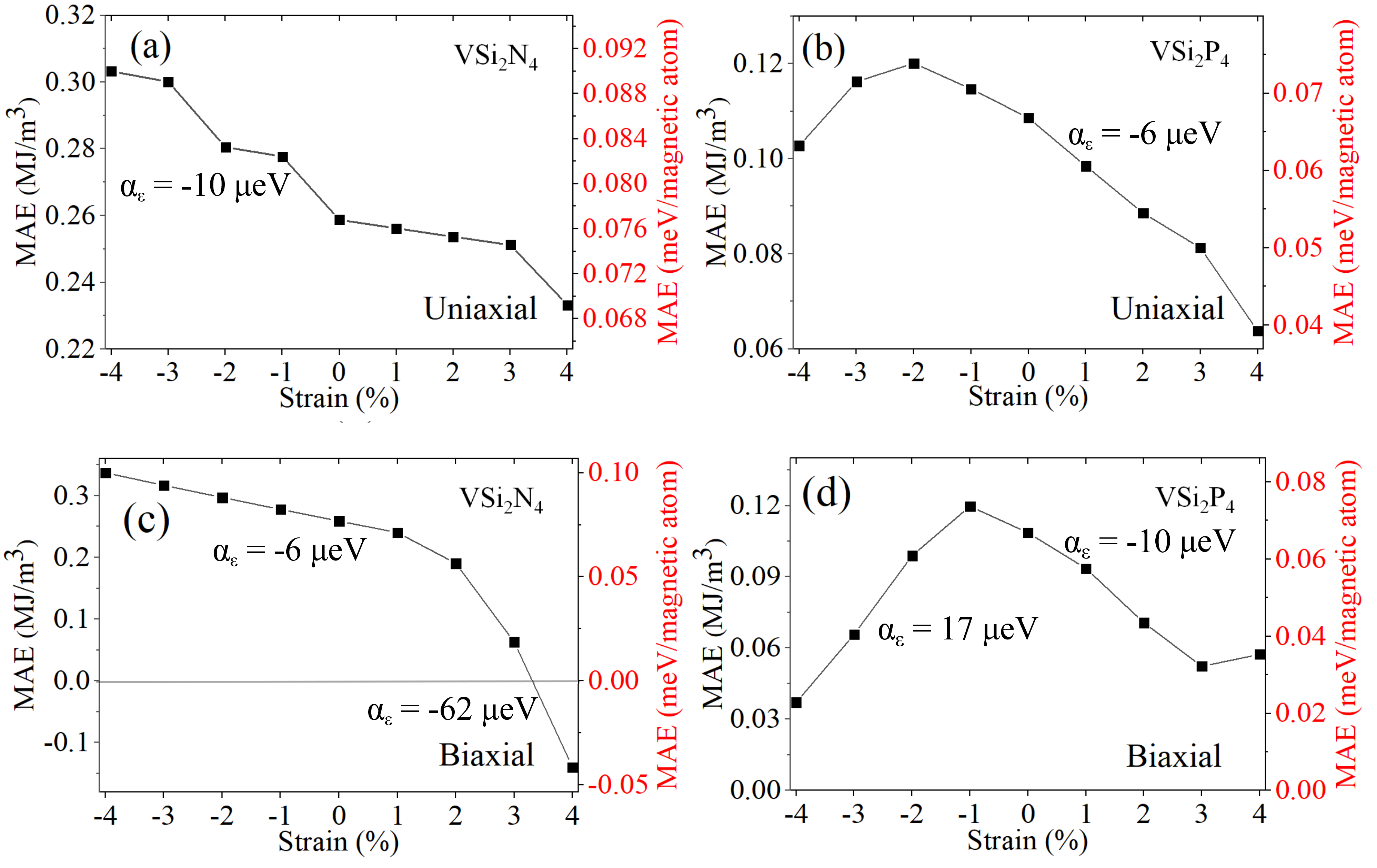}
\caption{MAE as a function of uniaxial and biaxial strain calculated with PBE(SOC)+U.
MAE versus uniaxial strain:
(a) VSi\textsubscript{2}N\textsubscript{4} and
(b) VSi\textsubscript{2}P\textsubscript{4}.
MAE versus biaxial strain:
(c) VSi\textsubscript{2}N\textsubscript{4} and
(d) VSi\textsubscript{2}P\textsubscript{4}.
Values are shown for the strain coefficients for different regions discussed in the text.
The orientation of the magnetization of \VN rotates from in-plane to out-of-plane with 3.3\%
biaxial strain.
}
\label{fig:strain}
\end{figure}

The equilibrium values of the MAE are listed in Table \ref{tab:mag_values_all}.
The equilibrium values of \VN and \VP range from 0.11 to 0.25 meV / magnetic atom.
For comparison,
monolayers of
CrCl\textsubscript{3},
CrBr\textsubscript{3},
and
CrI\textsubscript{3}
have equilibrium
MAEs of
$0.02$ meV,
$0.16$ meV,
and
$0.8$ meV,
respectively.\cite{MAECrX3}
Two-dimensional
FeCl\textsubscript{2},
NiI\textsubscript{2},
Fe\textsubscript{3}P,
and
Fe\textsubscript{3}GeTe\textsubscript{2}
have equilibrium MAE values of
$0.07$ meV,
$0.11$ meV,
$0.72$ meV,
and
$1$ meV,
respectively in their monolayer limit.\cite{MAEFe3GeTe2,MAEFeCl2,MAEFe3P,MAENiI2}
In general, the MAE values of \VN, and \VP are on the lower end of values for other 2D magnetic
materials.
%

%-------------------------  MAE vs. Strain  --------------------------------------------
The calculated values of the MAE of \VN and \VP as a function of uniaxial strain
and biaxial strain are shown in Fig.~\ref{fig:strain}.
We quantify the sensitivity by defining a strain coefficient as
$\alpha_\ep = dE_{\rm MAE}/d\ep$.
For small strain of both types in both materials,
the values for $\alpha_\ep$ are well below the value of 32 $\mu$eV/\%strain
recently calculated for a 1.1 nm slab of CrSb.\cite{IJPark_CrSb_PRB20}
For compressive uniaxial and biaxial strain in \VN,
$\alpha_\ep = -10$ and $-6$ $\mu$eV/\%strain, respectively.
For \VP,
$\alpha_\ep = -6$ $\mu$eV/\%strain for uniaxial tensile strain,
-10  $\mu$eV/\%strain for biaxial tensile strain, and
17 $\mu$eV/\%strain for biaxial compressive strain.
For large (3-4\%) biaxial strain in \VN, the magnitude of the sensitivity increases.
At a strain of 3.3\%, the MAE changes sign, and the spins rotate from in-plane to
out-of-plane.
This magnitude of strain should be experimentally accessible, since 2D materials
can sustain large strain.\cite{HighStrain}
%*************** MAE versus filling *************************
\begin{figure}
\includegraphics[width=\linewidth]{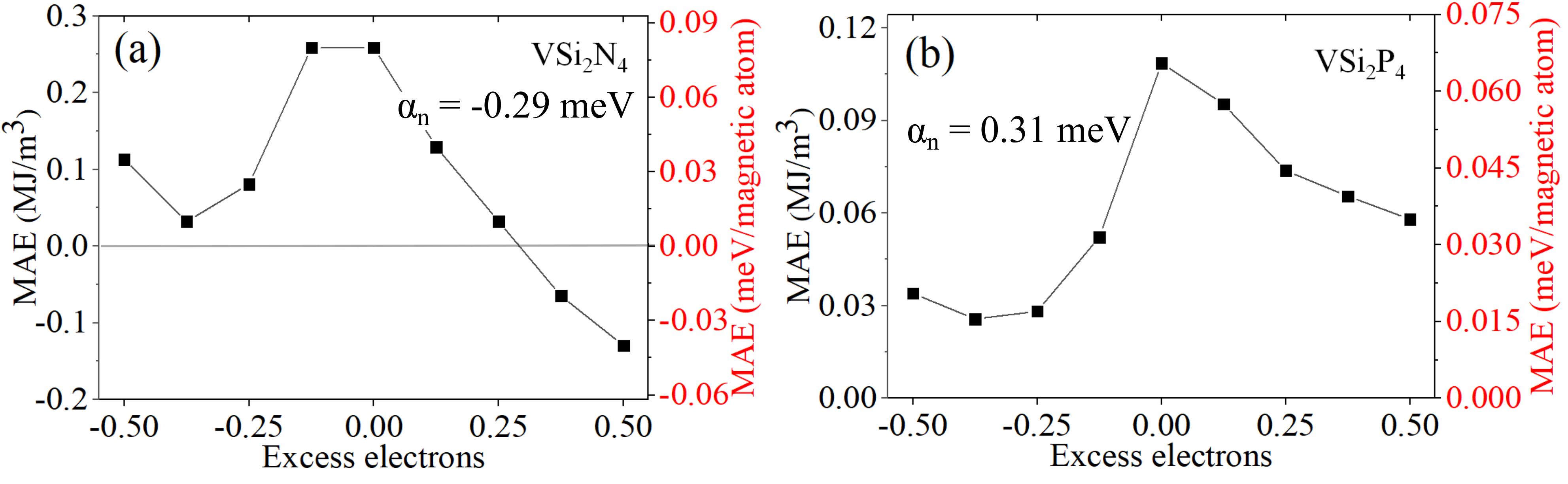}
\caption{MAE as a function of band filling calculated with PBE(SOC)+U.
MAE as a function of excess electrons per unit cell for
(a) VSi\textsubscript{2}N\textsubscript{4} and
(b) VSi\textsubscript{2}P\textsubscript{4}.
Values are shown for the filling coefficients discussed in the text.
The orientation of the magnetization of \VN rotates from in-plane to out-of-plane with electron doping
of $2.0\times 10^{14}$ cm$^{-2}$.
}
\label{fig:Filling}
\end{figure}

The effects of band filling on the MAE are shown in Fig.~\ref{fig:Filling}.
The MAE is reduced in \VN and \VP for both electron and hole doing.
To quantify the sensitivity of the MAE to charge filling
we define the parameter\cite{IJPark_CrSb_PRB20}
$\alpha_{n} = dE_{\rm MAE} / dn$ evaluated at zero filling.
For electron filling in \VN, $\alpha_n = -0.29$ meV, and for hole filling in \VP, $\alpha_n = 0.31$ meV.
What is most interesting is that the sign of the MAE can be reversed in \VN at an electron filling
of 0.28 electrons per unit cell corresponding to a sheet
carrier concentration of $n_s = 2.0 \times 10^{14}$ cm$^{-2}$.
At this doping density, the orientation of the magnetization of \VN rotates from in-plane to out-of-plane.
These densities are high, however since the thickness of the monolayer is under 1 nm, such densities could be
experimentally accessible using electrolytic gating.

In conclusion, \NbN, \VN, and \VP are easy-plane, semiconducting FMs with negligible in-plane anisotropy.
Exchange constants extracted from PBE+U DFT used in MC calculations predict Curie temperatures for
monolayer \VP and \VN of 350 K and 452 K, respectively.
The effect of the increasing levels of theory, PBE, PBE+U, and HSE06, is to successively increase the
energy gap between the two spin-polarized d$_{z^2}$ bands, so that at the HSE06 level, the two bands are completely
gapped which maximizes the spin polarization, magnetic moment, and the exchange constant.
The predicted $T_C$ increases with the increasing gap, giving a maximum value of 506 K for \VN
using the exchange constant determined from the HSE06 calculation.
Since \VN has the same structure, surface chemistry, and formation enthalpy as \MoN, it is expected to also be
air and water stable.
The magnetic anisotropy energies of \VN and \VP,
determined by the difference in the total energies resulting from out-of-plane versus in-plane alignment of the
magnetic moments,
are small, ranging from 0.11 to 0.25 meV / magnetic atom.
Tensile biaxial strain of 3.3\% in \VN causes the MAE to change sign such that the magnetic moments
rotate from in-plane to out-of-plane.
Band filling of 0.28 electrons per unit cell ($2.0 \times 10^{14}$ cm$^{-2}$)
also causes \VN to switch from in-plane to perpendicular magnetic anisotropy.

\noindent
{\em Supplementary Material}:
The supplementary material contains plots of orbital projected band diagrams,
comparisons of electronic bandstructures calculated with PBE, PBE+U, and HSE06 functionals,
and MC calculations of the normalized magnetization as a function of temperature with
exchange constants extracted from PBE DFT.
It also contains an extended comparison of Curie temperatures of 2D materials.

\noindent
{\em Acknowledgement}:
This work used the Extreme Science and Engineering Discovery Environment (XSEDE),\cite{towns2014xsede}
which is supported by National Science Foundation Grant No. ACI-1548562 and allocation ID TG-DMR130081.
MRKA thanks A. Bafekry from University of Guilan, and San-Dong Guo from Xi'an
University of Posts and Telecommunications for providing initial crystal structures.

\noindent
{\em Data Availability Statement}:
The data that support the findings of this study are available within the article and its supplementary material.

\bibliography{references}% Produces the bibliography via BibTeX.

%apsrev4-2.bst 2019-01-14 (MD) hand-edited version of apsrev4-1.bst
%Control: key (0)
%Control: author (72) initials jnrlst
%Control: editor formatted (1) identically to author
%Control: production of article title (-1) disabled
%Control: page (0) single
%Control: year (1) truncated
%Control: production of eprint (0) enabled
\providecommand{\noopsort}[1]{}\providecommand{\singleletter}[1]{#1}%
\begin{thebibliography}{64}%
\makeatletter
\providecommand \@ifxundefined [1]{%
 \@ifx{#1\undefined}
}%
\providecommand \@ifnum [1]{%
 \ifnum #1\expandafter \@firstoftwo
 \else \expandafter \@secondoftwo
 \fi
}%
\providecommand \@ifx [1]{%
 \ifx #1\expandafter \@firstoftwo
 \else \expandafter \@secondoftwo
 \fi
}%
\providecommand \natexlab [1]{#1}%
\providecommand \enquote  [1]{``#1''}%
\providecommand \bibnamefont  [1]{#1}%
\providecommand \bibfnamefont [1]{#1}%
\providecommand \citenamefont [1]{#1}%
\providecommand \href@noop [0]{\@secondoftwo}%
\providecommand \href [0]{\begingroup \@sanitize@url \@href}%
\providecommand \@href[1]{\@@startlink{#1}\@@href}%
\providecommand \@@href[1]{\endgroup#1\@@endlink}%
\providecommand \@sanitize@url [0]{\catcode `\\12\catcode `\$12\catcode
  `\&12\catcode `\#12\catcode `\^12\catcode `\_12\catcode `\%12\relax}%
\providecommand \@@startlink[1]{}%
\providecommand \@@endlink[0]{}%
\providecommand \url  [0]{\begingroup\@sanitize@url \@url }%
\providecommand \@url [1]{\endgroup\@href {#1}{\urlprefix }}%
\providecommand \urlprefix  [0]{URL }%
\providecommand \Eprint [0]{\href }%
\providecommand \doibase [0]{https://doi.org/}%
\providecommand \selectlanguage [0]{\@gobble}%
\providecommand \bibinfo  [0]{\@secondoftwo}%
\providecommand \bibfield  [0]{\@secondoftwo}%
\providecommand \translation [1]{[#1]}%
\providecommand \BibitemOpen [0]{}%
\providecommand \bibitemStop [0]{}%
\providecommand \bibitemNoStop [0]{.\EOS\space}%
\providecommand \EOS [0]{\spacefactor3000\relax}%
\providecommand \BibitemShut  [1]{\csname bibitem#1\endcsname}%
\let\auto@bib@innerbib\@empty
%</preamble>
\bibitem [{\citenamefont {Friend}\ and\ \citenamefont {Yoffe}(1987)}]{Yoffe87}%
  \BibitemOpen
  \bibfield  {author} {\bibinfo {author} {\bibfnamefont {R.~H.}\ \bibnamefont
  {Friend}}\ and\ \bibinfo {author} {\bibfnamefont {A.~D.}\ \bibnamefont
  {Yoffe}},\ }\href@noop {} {\bibfield  {journal} {\bibinfo  {journal} {Adv.
  Phys.}\ }\textbf {\bibinfo {volume} {36}},\ \bibinfo {pages} {1} (\bibinfo
  {year} {1987})}\BibitemShut {NoStop}%
\bibitem [{\citenamefont {Klemm}(2012)}]{Klemm_Layered_SCs}%
  \BibitemOpen
  \bibfield  {author} {\bibinfo {author} {\bibfnamefont {R.~A.}\ \bibnamefont
  {Klemm}},\ }\href@noop {} {\emph {\bibinfo {title} {Layered
  Superconductors}}},\ Vol.~\bibinfo {volume} {1}\ (\bibinfo  {publisher}
  {Oxford University Press},\ \bibinfo {address} {New York},\ \bibinfo {year}
  {2012})\BibitemShut {NoStop}%
\bibitem [{\citenamefont {de~Jongh}(1989)}]{Jongh_Mag_Props_Layered_Mats}%
  \BibitemOpen
  \bibinfo {editor} {\bibfnamefont {L.~J.}\ \bibnamefont {de~Jongh}},\ ed.,\
  \href@noop {} {\emph {\bibinfo {title} {Magnetic Properties of Layered
  Transition Metal Compounds}}},\ \bibinfo {series} {Physics and Chemistry of
  Materials with Low-Dimensional Structures}, Vol.~\bibinfo {volume} {9}\
  (\bibinfo  {publisher} {Kluwer Academic Publishers},\ \bibinfo {year}
  {1989})\BibitemShut {NoStop}%
\bibitem [{\citenamefont {Radisavljevic}\ \emph {et~al.}(2011)\citenamefont
  {Radisavljevic}, \citenamefont {Radenovic}, \citenamefont {Brivio},
  \citenamefont {Giacometti},\ and\ \citenamefont
  {Kis}}]{2011_ML_MoS2_FET_AKis}%
  \BibitemOpen
  \bibfield  {author} {\bibinfo {author} {\bibfnamefont {B.}~\bibnamefont
  {Radisavljevic}}, \bibinfo {author} {\bibfnamefont {A.}~\bibnamefont
  {Radenovic}}, \bibinfo {author} {\bibfnamefont {J.}~\bibnamefont {Brivio}},
  \bibinfo {author} {\bibfnamefont {V.}~\bibnamefont {Giacometti}},\ and\
  \bibinfo {author} {\bibfnamefont {A.}~\bibnamefont {Kis}},\ }\href
  {https://doi.org/10.1038/nnano.2010.279} {\bibfield  {journal} {\bibinfo
  {journal} {Nature nanotechnology}\ }\textbf {\bibinfo {volume} {6}},\
  \bibinfo {pages} {147} (\bibinfo {year} {2011})}\BibitemShut {NoStop}%
\bibitem [{\citenamefont {Kang}\ \emph {et~al.}(2020)\citenamefont {Kang},
  \citenamefont {Lee}, \citenamefont {Kim}, \citenamefont {Capasso},
  \citenamefont {Kang}, \citenamefont {Park}, \citenamefont {Lee},\ and\
  \citenamefont {Lee}}]{2020_2D_Review_2DMat}%
  \BibitemOpen
  \bibfield  {author} {\bibinfo {author} {\bibfnamefont {S.}~\bibnamefont
  {Kang}}, \bibinfo {author} {\bibfnamefont {D.}~\bibnamefont {Lee}}, \bibinfo
  {author} {\bibfnamefont {J.}~\bibnamefont {Kim}}, \bibinfo {author}
  {\bibfnamefont {A.}~\bibnamefont {Capasso}}, \bibinfo {author} {\bibfnamefont
  {H.~S.}\ \bibnamefont {Kang}}, \bibinfo {author} {\bibfnamefont {J.-W.}\
  \bibnamefont {Park}}, \bibinfo {author} {\bibfnamefont {C.-H.}\ \bibnamefont
  {Lee}},\ and\ \bibinfo {author} {\bibfnamefont {G.-H.}\ \bibnamefont {Lee}},\
  }\href {https://doi.org/10.1088/2053-1583/ab6267} {\bibfield  {journal}
  {\bibinfo  {journal} {2D Materials}\ }\textbf {\bibinfo {volume} {7}},\
  \bibinfo {pages} {022003} (\bibinfo {year} {2020})}\BibitemShut {NoStop}%
\bibitem [{\citenamefont {Wang}\ \emph {et~al.}(2012)\citenamefont {Wang},
  \citenamefont {Kalantar-Zadeh}, \citenamefont {Kis}, \citenamefont
  {Coleman},\ and\ \citenamefont {Strano}}]{wang2012electronics}%
  \BibitemOpen
  \bibfield  {author} {\bibinfo {author} {\bibfnamefont {Q.~H.}\ \bibnamefont
  {Wang}}, \bibinfo {author} {\bibfnamefont {K.}~\bibnamefont
  {Kalantar-Zadeh}}, \bibinfo {author} {\bibfnamefont {A.}~\bibnamefont {Kis}},
  \bibinfo {author} {\bibfnamefont {J.~N.}\ \bibnamefont {Coleman}},\ and\
  \bibinfo {author} {\bibfnamefont {M.~S.}\ \bibnamefont {Strano}},\ }\href
  {https://doi.org/10.1038/nnano.2012.193} {\bibfield  {journal} {\bibinfo
  {journal} {Nature nanotechnology}\ }\textbf {\bibinfo {volume} {7}},\
  \bibinfo {pages} {699} (\bibinfo {year} {2012})}\BibitemShut {NoStop}%
\bibitem [{\citenamefont {Fiori}\ \emph {et~al.}(2014)\citenamefont {Fiori},
  \citenamefont {Bonaccorso}, \citenamefont {Iannaccone}, \citenamefont
  {Palacios}, \citenamefont {Neumaier}, \citenamefont {Seabaugh}, \citenamefont
  {Banerjee},\ and\ \citenamefont {Colombo}}]{Fiori_2DMat_Review_NNano14}%
  \BibitemOpen
  \bibfield  {author} {\bibinfo {author} {\bibfnamefont {G.}~\bibnamefont
  {Fiori}}, \bibinfo {author} {\bibfnamefont {F.}~\bibnamefont {Bonaccorso}},
  \bibinfo {author} {\bibfnamefont {G.}~\bibnamefont {Iannaccone}}, \bibinfo
  {author} {\bibfnamefont {T.}~\bibnamefont {Palacios}}, \bibinfo {author}
  {\bibfnamefont {D.}~\bibnamefont {Neumaier}}, \bibinfo {author}
  {\bibfnamefont {A.}~\bibnamefont {Seabaugh}}, \bibinfo {author}
  {\bibfnamefont {S.~K.}\ \bibnamefont {Banerjee}},\ and\ \bibinfo {author}
  {\bibfnamefont {L.}~\bibnamefont {Colombo}},\ }\href
  {https://doi.org/10.1038/nnano.2014.207} {\bibfield  {journal} {\bibinfo
  {journal} {Nature Nanotechnology}\ }\textbf {\bibinfo {volume} {9}},\
  \bibinfo {pages} {768} (\bibinfo {year} {2014})}\BibitemShut {NoStop}%
\bibitem [{\citenamefont {Li}\ \emph {et~al.}(2021)\citenamefont {Li},
  \citenamefont {Qian},\ and\ \citenamefont
  {Li}}]{2021_Phase_Transitions_NatRev}%
  \BibitemOpen
  \bibfield  {author} {\bibinfo {author} {\bibfnamefont {W.}~\bibnamefont
  {Li}}, \bibinfo {author} {\bibfnamefont {X.}~\bibnamefont {Qian}},\ and\
  \bibinfo {author} {\bibfnamefont {J.}~\bibnamefont {Li}},\ }\href
  {https://doi.org/10.1038/s41578-021-00304-0} {\bibfield  {journal} {\bibinfo
  {journal} {Nature Reviews Materials}\ ,\ \bibinfo {pages} {1}} (\bibinfo
  {year} {2021})}\BibitemShut {NoStop}%
\bibitem [{\citenamefont {Huang}\ \emph
  {et~al.}(2017{\natexlab{a}})\citenamefont {Huang}, \citenamefont {Clark},
  \citenamefont {Navarro-Moratalla}, \citenamefont {Klein}, \citenamefont
  {Cheng}, \citenamefont {Seyler}, \citenamefont {Zhong}, \citenamefont
  {Schmidgall}, \citenamefont {McGuire}, \citenamefont {Cobden}, \citenamefont
  {Yao}, \citenamefont {Xiao}, \citenamefont {Jarillo-Herrero},\ and\
  \citenamefont {Xu}}]{CrI3}%
  \BibitemOpen
  \bibfield  {author} {\bibinfo {author} {\bibfnamefont {B.}~\bibnamefont
  {Huang}}, \bibinfo {author} {\bibfnamefont {G.}~\bibnamefont {Clark}},
  \bibinfo {author} {\bibfnamefont {E.}~\bibnamefont {Navarro-Moratalla}},
  \bibinfo {author} {\bibfnamefont {D.~R.}\ \bibnamefont {Klein}}, \bibinfo
  {author} {\bibfnamefont {R.}~\bibnamefont {Cheng}}, \bibinfo {author}
  {\bibfnamefont {K.~L.}\ \bibnamefont {Seyler}}, \bibinfo {author}
  {\bibfnamefont {D.}~\bibnamefont {Zhong}}, \bibinfo {author} {\bibfnamefont
  {E.}~\bibnamefont {Schmidgall}}, \bibinfo {author} {\bibfnamefont {M.~A.}\
  \bibnamefont {McGuire}}, \bibinfo {author} {\bibfnamefont {D.~H.}\
  \bibnamefont {Cobden}}, \bibinfo {author} {\bibfnamefont {W.}~\bibnamefont
  {Yao}}, \bibinfo {author} {\bibfnamefont {D.}~\bibnamefont {Xiao}}, \bibinfo
  {author} {\bibfnamefont {P.}~\bibnamefont {Jarillo-Herrero}},\ and\ \bibinfo
  {author} {\bibfnamefont {X.}~\bibnamefont {Xu}},\ }\href
  {https://doi.org/10.1038/nature22391} {\bibfield  {journal} {\bibinfo
  {journal} {Nature}\ }\textbf {\bibinfo {volume} {546}},\ \bibinfo {pages}
  {270} (\bibinfo {year} {2017}{\natexlab{a}})}\BibitemShut {NoStop}%
\bibitem [{\citenamefont {Gong}\ \emph {et~al.}(2017)\citenamefont {Gong},
  \citenamefont {Li}, \citenamefont {Li}, \citenamefont {Ji}, \citenamefont
  {Stern}, \citenamefont {Xia}, \citenamefont {Cao}, \citenamefont {Bao},
  \citenamefont {Wang}, \citenamefont {Wang}, \citenamefont {Qiu},
  \citenamefont {Cava}, \citenamefont {Louie}, \citenamefont {Xia},\ and\
  \citenamefont {Zhang}}]{2017_CGT_Nat}%
  \BibitemOpen
  \bibfield  {author} {\bibinfo {author} {\bibfnamefont {C.}~\bibnamefont
  {Gong}}, \bibinfo {author} {\bibfnamefont {L.}~\bibnamefont {Li}}, \bibinfo
  {author} {\bibfnamefont {Z.}~\bibnamefont {Li}}, \bibinfo {author}
  {\bibfnamefont {H.}~\bibnamefont {Ji}}, \bibinfo {author} {\bibfnamefont
  {A.}~\bibnamefont {Stern}}, \bibinfo {author} {\bibfnamefont
  {Y.}~\bibnamefont {Xia}}, \bibinfo {author} {\bibfnamefont {T.}~\bibnamefont
  {Cao}}, \bibinfo {author} {\bibfnamefont {W.}~\bibnamefont {Bao}}, \bibinfo
  {author} {\bibfnamefont {C.}~\bibnamefont {Wang}}, \bibinfo {author}
  {\bibfnamefont {Y.}~\bibnamefont {Wang}}, \bibinfo {author} {\bibfnamefont
  {Z.~Q.}\ \bibnamefont {Qiu}}, \bibinfo {author} {\bibfnamefont {R.~J.}\
  \bibnamefont {Cava}}, \bibinfo {author} {\bibfnamefont {S.~G.}\ \bibnamefont
  {Louie}}, \bibinfo {author} {\bibfnamefont {J.}~\bibnamefont {Xia}},\ and\
  \bibinfo {author} {\bibfnamefont {X.}~\bibnamefont {Zhang}},\ }\href
  {https://doi.org/10.1038/nature22060} {\bibfield  {journal} {\bibinfo
  {journal} {Nature}\ }\textbf {\bibinfo {volume} {546}},\ \bibinfo {pages}
  {265} (\bibinfo {year} {2017})}\BibitemShut {NoStop}%
\bibitem [{\citenamefont {Deng}\ \emph {et~al.}(2018)\citenamefont {Deng},
  \citenamefont {Yu}, \citenamefont {Song}, \citenamefont {Zhang},
  \citenamefont {Wang}, \citenamefont {Sun}, \citenamefont {Yi}, \citenamefont
  {Wu}, \citenamefont {Wu}, \citenamefont {Zhu}, \citenamefont {Wang},
  \citenamefont {Chen},\ and\ \citenamefont
  {Zhang}}]{2018_Fe3GeTe2_FM_Gate_Tunable_Nat}%
  \BibitemOpen
  \bibfield  {author} {\bibinfo {author} {\bibfnamefont {Y.}~\bibnamefont
  {Deng}}, \bibinfo {author} {\bibfnamefont {Y.}~\bibnamefont {Yu}}, \bibinfo
  {author} {\bibfnamefont {Y.}~\bibnamefont {Song}}, \bibinfo {author}
  {\bibfnamefont {J.}~\bibnamefont {Zhang}}, \bibinfo {author} {\bibfnamefont
  {N.~Z.}\ \bibnamefont {Wang}}, \bibinfo {author} {\bibfnamefont
  {Z.}~\bibnamefont {Sun}}, \bibinfo {author} {\bibfnamefont {Y.}~\bibnamefont
  {Yi}}, \bibinfo {author} {\bibfnamefont {Y.~Z.}\ \bibnamefont {Wu}}, \bibinfo
  {author} {\bibfnamefont {S.}~\bibnamefont {Wu}}, \bibinfo {author}
  {\bibfnamefont {J.}~\bibnamefont {Zhu}}, \bibinfo {author} {\bibfnamefont
  {J.}~\bibnamefont {Wang}}, \bibinfo {author} {\bibfnamefont {X.~H.}\
  \bibnamefont {Chen}},\ and\ \bibinfo {author} {\bibfnamefont
  {Y.}~\bibnamefont {Zhang}},\ }\href
  {https://doi.org/10.1038/s41586-018-0626-9} {\bibfield  {journal} {\bibinfo
  {journal} {Nature}\ }\textbf {\bibinfo {volume} {563}},\ \bibinfo {pages}
  {94} (\bibinfo {year} {2018})}\BibitemShut {NoStop}%
\bibitem [{\citenamefont {Gibertini}\ \emph {et~al.}(2019)\citenamefont
  {Gibertini}, \citenamefont {Koperski}, \citenamefont {Morpurgo},\ and\
  \citenamefont {Novoselov}}]{2019_2D_Mag_Novoselov_Rev_NMat}%
  \BibitemOpen
  \bibfield  {author} {\bibinfo {author} {\bibfnamefont {M.}~\bibnamefont
  {Gibertini}}, \bibinfo {author} {\bibfnamefont {M.}~\bibnamefont {Koperski}},
  \bibinfo {author} {\bibfnamefont {A.~F.}\ \bibnamefont {Morpurgo}},\ and\
  \bibinfo {author} {\bibfnamefont {K.~S.}\ \bibnamefont {Novoselov}},\ }\href
  {https://doi.org/10.1038/s41565-019-0438-6} {\bibfield  {journal} {\bibinfo
  {journal} {Nature Nanotechnology}\ }\textbf {\bibinfo {volume} {14}},\
  \bibinfo {pages} {408} (\bibinfo {year} {2019})}\BibitemShut {NoStop}%
\bibitem [{\citenamefont {Li}\ \emph {et~al.}(2019)\citenamefont {Li},
  \citenamefont {Ruan},\ and\ \citenamefont {Zeng}}]{2019_2D_Mag_Rev_AdvMat}%
  \BibitemOpen
  \bibfield  {author} {\bibinfo {author} {\bibfnamefont {H.}~\bibnamefont
  {Li}}, \bibinfo {author} {\bibfnamefont {S.}~\bibnamefont {Ruan}},\ and\
  \bibinfo {author} {\bibfnamefont {Y.-J.}\ \bibnamefont {Zeng}},\ }\href
  {https://doi.org/10.1002/adma.201900065} {\bibfield  {journal} {\bibinfo
  {journal} {Advanced Materials}\ }\textbf {\bibinfo {volume} {31}},\ \bibinfo
  {pages} {1900065} (\bibinfo {year} {2019})}\BibitemShut {NoStop}%
\bibitem [{\citenamefont {Chittari}\ \emph {et~al.}(2020)\citenamefont
  {Chittari}, \citenamefont {Lee}, \citenamefont {Banerjee}, \citenamefont
  {MacDonald}, \citenamefont {Hwang},\ and\ \citenamefont
  {Jung}}]{2020_MAX3_FM_MacDonald_PRB}%
  \BibitemOpen
  \bibfield  {author} {\bibinfo {author} {\bibfnamefont {B.~L.}\ \bibnamefont
  {Chittari}}, \bibinfo {author} {\bibfnamefont {D.}~\bibnamefont {Lee}},
  \bibinfo {author} {\bibfnamefont {N.}~\bibnamefont {Banerjee}}, \bibinfo
  {author} {\bibfnamefont {A.~H.}\ \bibnamefont {MacDonald}}, \bibinfo {author}
  {\bibfnamefont {E.}~\bibnamefont {Hwang}},\ and\ \bibinfo {author}
  {\bibfnamefont {J.}~\bibnamefont {Jung}},\ }\href
  {https://doi.org/10.1103/PhysRevB.101.085415} {\bibfield  {journal} {\bibinfo
   {journal} {Phys. Rev. B}\ }\textbf {\bibinfo {volume} {101}},\ \bibinfo
  {pages} {085415} (\bibinfo {year} {2020})}\BibitemShut {NoStop}%
\bibitem [{\citenamefont {Kabiraj}\ \emph {et~al.}(2020)\citenamefont
  {Kabiraj}, \citenamefont {Kumar},\ and\ \citenamefont
  {Mahapatra}}]{2020_Hi_Thruput_2D_FM_NPJCM}%
  \BibitemOpen
  \bibfield  {author} {\bibinfo {author} {\bibfnamefont {A.}~\bibnamefont
  {Kabiraj}}, \bibinfo {author} {\bibfnamefont {M.}~\bibnamefont {Kumar}},\
  and\ \bibinfo {author} {\bibfnamefont {S.}~\bibnamefont {Mahapatra}},\ }\href
  {https://doi.org/10.1038/s41524-020-0300-2} {\bibfield  {journal} {\bibinfo
  {journal} {npj Computational Materials}\ }\textbf {\bibinfo {volume} {6}},\
  \bibinfo {pages} {35} (\bibinfo {year} {2020})}\BibitemShut {NoStop}%
\bibitem [{\citenamefont {Han}\ \emph {et~al.}(2020{\natexlab{a}})\citenamefont
  {Han}, \citenamefont {Jiang},\ and\ \citenamefont
  {Yan}}]{2020_CrSeX_FM_JPCehmC}%
  \BibitemOpen
  \bibfield  {author} {\bibinfo {author} {\bibfnamefont {R.}~\bibnamefont
  {Han}}, \bibinfo {author} {\bibfnamefont {Z.}~\bibnamefont {Jiang}},\ and\
  \bibinfo {author} {\bibfnamefont {Y.}~\bibnamefont {Yan}},\ }\href
  {https://doi.org/10.1021/acs.jpcc.0c01307} {\bibfield  {journal} {\bibinfo
  {journal} {The Journal of Physical Chemistry C}\ }\textbf {\bibinfo {volume}
  {124}},\ \bibinfo {pages} {7956} (\bibinfo {year}
  {2020}{\natexlab{a}})}\BibitemShut {NoStop}%
\bibitem [{\citenamefont {Jothi}\ \emph {et~al.}(2020)\citenamefont {Jothi},
  \citenamefont {Scheifers}, \citenamefont {Zhang}, \citenamefont {Alghamdi},
  \citenamefont {Stekovic}, \citenamefont {Itkis}, \citenamefont {Shi},\ and\
  \citenamefont {Fokwa}}]{2020_FGT_PSS_Fokwa}%
  \BibitemOpen
  \bibfield  {author} {\bibinfo {author} {\bibfnamefont {P.~R.}\ \bibnamefont
  {Jothi}}, \bibinfo {author} {\bibfnamefont {J.~P.}\ \bibnamefont
  {Scheifers}}, \bibinfo {author} {\bibfnamefont {Y.}~\bibnamefont {Zhang}},
  \bibinfo {author} {\bibfnamefont {M.}~\bibnamefont {Alghamdi}}, \bibinfo
  {author} {\bibfnamefont {D.}~\bibnamefont {Stekovic}}, \bibinfo {author}
  {\bibfnamefont {M.~E.}\ \bibnamefont {Itkis}}, \bibinfo {author}
  {\bibfnamefont {J.}~\bibnamefont {Shi}},\ and\ \bibinfo {author}
  {\bibfnamefont {B.~P.~T.}\ \bibnamefont {Fokwa}},\ }\href
  {https://doi.org/10.1002/pssr.201900666} {\bibfield  {journal} {\bibinfo
  {journal} {Physica Status Solidi (RRL) - Rapid Research Letters}\ }\textbf
  {\bibinfo {volume} {14}},\ \bibinfo {pages} {1900666} (\bibinfo {year}
  {2020})}\BibitemShut {NoStop}%
\bibitem [{\citenamefont {Khan}\ \emph {et~al.}(2020)\citenamefont {Khan},
  \citenamefont {Obaidulla}, \citenamefont {Habib}, \citenamefont {Gayen},
  \citenamefont {Liang}, \citenamefont {Wang},\ and\ \citenamefont
  {Xu}}]{2020_2D_Mag_Mats_NanoToday}%
  \BibitemOpen
  \bibfield  {author} {\bibinfo {author} {\bibfnamefont {Y.}~\bibnamefont
  {Khan}}, \bibinfo {author} {\bibfnamefont {S.~M.}\ \bibnamefont {Obaidulla}},
  \bibinfo {author} {\bibfnamefont {M.~R.}\ \bibnamefont {Habib}}, \bibinfo
  {author} {\bibfnamefont {A.}~\bibnamefont {Gayen}}, \bibinfo {author}
  {\bibfnamefont {T.}~\bibnamefont {Liang}}, \bibinfo {author} {\bibfnamefont
  {X.}~\bibnamefont {Wang}},\ and\ \bibinfo {author} {\bibfnamefont
  {M.}~\bibnamefont {Xu}},\ }\href
  {https://doi.org/10.1016/j.nantod.2020.100902} {\bibfield  {journal}
  {\bibinfo  {journal} {Nano Today}\ }\textbf {\bibinfo {volume} {34}},\
  \bibinfo {pages} {100902} (\bibinfo {year} {2020})}\BibitemShut {NoStop}%
\bibitem [{\citenamefont {Hong}\ \emph {et~al.}(2020)\citenamefont {Hong},
  \citenamefont {Liu}, \citenamefont {Wang}, \citenamefont {Zhou},
  \citenamefont {Ma}, \citenamefont {Xu}, \citenamefont {Feng}, \citenamefont
  {Chen}, \citenamefont {Chen}, \citenamefont {Sun}, \citenamefont {Chen},
  \citenamefont {Cheng},\ and\ \citenamefont {Ren}}]{2020_MSi2N4_Sci}%
  \BibitemOpen
  \bibfield  {author} {\bibinfo {author} {\bibfnamefont {Y.-L.}\ \bibnamefont
  {Hong}}, \bibinfo {author} {\bibfnamefont {Z.}~\bibnamefont {Liu}}, \bibinfo
  {author} {\bibfnamefont {L.}~\bibnamefont {Wang}}, \bibinfo {author}
  {\bibfnamefont {T.}~\bibnamefont {Zhou}}, \bibinfo {author} {\bibfnamefont
  {W.}~\bibnamefont {Ma}}, \bibinfo {author} {\bibfnamefont {C.}~\bibnamefont
  {Xu}}, \bibinfo {author} {\bibfnamefont {S.}~\bibnamefont {Feng}}, \bibinfo
  {author} {\bibfnamefont {L.}~\bibnamefont {Chen}}, \bibinfo {author}
  {\bibfnamefont {M.-L.}\ \bibnamefont {Chen}}, \bibinfo {author}
  {\bibfnamefont {D.-M.}\ \bibnamefont {Sun}}, \bibinfo {author} {\bibfnamefont
  {X.-Q.}\ \bibnamefont {Chen}}, \bibinfo {author} {\bibfnamefont {H.-M.}\
  \bibnamefont {Cheng}},\ and\ \bibinfo {author} {\bibfnamefont
  {W.}~\bibnamefont {Ren}},\ }\href {https://doi.org/10.1126/science.abb7023}
  {\bibfield  {journal} {\bibinfo  {journal} {Science}\ }\textbf {\bibinfo
  {volume} {369}},\ \bibinfo {pages} {670} (\bibinfo {year}
  {2020})}\BibitemShut {NoStop}%
\bibitem [{\citenamefont {Gillgren}\ \emph {et~al.}(2014)\citenamefont
  {Gillgren}, \citenamefont {Wickramaratne}, \citenamefont {Shi}, \citenamefont
  {Espiritu}, \citenamefont {Yang}, \citenamefont {Hu}, \citenamefont {Wei},
  \citenamefont {Liu}, \citenamefont {Mao}, \citenamefont {Watanabe},
  \citenamefont {Taniguchi}, \citenamefont {Bockrath}, \citenamefont {Barlas},
  \citenamefont {Lake},\ and\ \citenamefont {Lau}}]{2014_BP_JLau_2DMats}%
  \BibitemOpen
  \bibfield  {author} {\bibinfo {author} {\bibfnamefont {N.}~\bibnamefont
  {Gillgren}}, \bibinfo {author} {\bibfnamefont {D.}~\bibnamefont
  {Wickramaratne}}, \bibinfo {author} {\bibfnamefont {Y.}~\bibnamefont {Shi}},
  \bibinfo {author} {\bibfnamefont {T.}~\bibnamefont {Espiritu}}, \bibinfo
  {author} {\bibfnamefont {J.}~\bibnamefont {Yang}}, \bibinfo {author}
  {\bibfnamefont {J.}~\bibnamefont {Hu}}, \bibinfo {author} {\bibfnamefont
  {J.}~\bibnamefont {Wei}}, \bibinfo {author} {\bibfnamefont {X.}~\bibnamefont
  {Liu}}, \bibinfo {author} {\bibfnamefont {Z.}~\bibnamefont {Mao}}, \bibinfo
  {author} {\bibfnamefont {K.}~\bibnamefont {Watanabe}}, \bibinfo {author}
  {\bibfnamefont {T.}~\bibnamefont {Taniguchi}}, \bibinfo {author}
  {\bibfnamefont {M.}~\bibnamefont {Bockrath}}, \bibinfo {author}
  {\bibfnamefont {Y.}~\bibnamefont {Barlas}}, \bibinfo {author} {\bibfnamefont
  {R.~K.}\ \bibnamefont {Lake}},\ and\ \bibinfo {author} {\bibfnamefont
  {C.~N.}\ \bibnamefont {Lau}},\ }\href
  {https://doi.org/10.1088/2053-1583/2/1/011001} {\bibfield  {journal}
  {\bibinfo  {journal} {2D Materials}\ }\textbf {\bibinfo {volume} {2}},\
  \bibinfo {pages} {011001} (\bibinfo {year} {2014})}\BibitemShut {NoStop}%
\bibitem [{\citenamefont {Guo}\ \emph {et~al.}(2020)\citenamefont {Guo},
  \citenamefont {Mu}, \citenamefont {Zhu},\ and\ \citenamefont
  {Chen}}]{2020_VSi2P4_FM_Piezo_PCCP}%
  \BibitemOpen
  \bibfield  {author} {\bibinfo {author} {\bibfnamefont {S.-D.}\ \bibnamefont
  {Guo}}, \bibinfo {author} {\bibfnamefont {W.-Q.}\ \bibnamefont {Mu}},
  \bibinfo {author} {\bibfnamefont {Y.-T.}\ \bibnamefont {Zhu}},\ and\ \bibinfo
  {author} {\bibfnamefont {X.-Q.}\ \bibnamefont {Chen}},\ }\href
  {https://doi.org/10.1039/D0CP05273F} {\bibfield  {journal} {\bibinfo
  {journal} {Phys. Chem. Chem. Phys.}\ }\textbf {\bibinfo {volume} {22}},\
  \bibinfo {pages} {28359} (\bibinfo {year} {2020})}\BibitemShut {NoStop}%
\bibitem [{\citenamefont {Cao}\ \emph {et~al.}(2021)\citenamefont {Cao},
  \citenamefont {Zhou}, \citenamefont {Wang}, \citenamefont {Ang},\ and\
  \citenamefont {Ang}}]{2020_MoSi2N4_Contacts_APL}%
  \BibitemOpen
  \bibfield  {author} {\bibinfo {author} {\bibfnamefont {L.}~\bibnamefont
  {Cao}}, \bibinfo {author} {\bibfnamefont {G.}~\bibnamefont {Zhou}}, \bibinfo
  {author} {\bibfnamefont {Q.}~\bibnamefont {Wang}}, \bibinfo {author}
  {\bibfnamefont {L.~K.}\ \bibnamefont {Ang}},\ and\ \bibinfo {author}
  {\bibfnamefont {Y.~S.}\ \bibnamefont {Ang}},\ }\href
  {https://doi.org/10.1063/5.0033241} {\bibfield  {journal} {\bibinfo
  {journal} {Applied Physics Letters}\ }\textbf {\bibinfo {volume} {118}},\
  \bibinfo {pages} {013106} (\bibinfo {year} {2021})}\BibitemShut {NoStop}%
\bibitem [{\citenamefont {Li}\ \emph {et~al.}(2020)\citenamefont {Li},
  \citenamefont {Wu}, \citenamefont {Feng}, \citenamefont {Guan}, \citenamefont
  {Feng}, \citenamefont {Yao},\ and\ \citenamefont
  {Yang}}]{2020_Valley_Dep_Properties_PRB}%
  \BibitemOpen
  \bibfield  {author} {\bibinfo {author} {\bibfnamefont {S.}~\bibnamefont
  {Li}}, \bibinfo {author} {\bibfnamefont {W.}~\bibnamefont {Wu}}, \bibinfo
  {author} {\bibfnamefont {X.}~\bibnamefont {Feng}}, \bibinfo {author}
  {\bibfnamefont {S.}~\bibnamefont {Guan}}, \bibinfo {author} {\bibfnamefont
  {W.}~\bibnamefont {Feng}}, \bibinfo {author} {\bibfnamefont {Y.}~\bibnamefont
  {Yao}},\ and\ \bibinfo {author} {\bibfnamefont {S.~A.}\ \bibnamefont
  {Yang}},\ }\href {https://doi.org/10.1103/PhysRevB.102.235435} {\bibfield
  {journal} {\bibinfo  {journal} {Phys. Rev. B}\ }\textbf {\bibinfo {volume}
  {102}},\ \bibinfo {pages} {235435} (\bibinfo {year} {2020})}\BibitemShut
  {NoStop}%
\bibitem [{\citenamefont {Wang}\ \emph {et~al.}(2021)\citenamefont {Wang},
  \citenamefont {Shi}, \citenamefont {Liu}, \citenamefont {Zhang},
  \citenamefont {Hong}, \citenamefont {Li}, \citenamefont {Gao}, \citenamefont
  {Chen}, \citenamefont {Ren}, \citenamefont {Cheng}, \citenamefont {Li},\ and\
  \citenamefont {Chen}}]{2021_MA2Z4_TI_to_Ising_SC_NComm}%
  \BibitemOpen
  \bibfield  {author} {\bibinfo {author} {\bibfnamefont {L.}~\bibnamefont
  {Wang}}, \bibinfo {author} {\bibfnamefont {Y.}~\bibnamefont {Shi}}, \bibinfo
  {author} {\bibfnamefont {M.}~\bibnamefont {Liu}}, \bibinfo {author}
  {\bibfnamefont {A.}~\bibnamefont {Zhang}}, \bibinfo {author} {\bibfnamefont
  {Y.-L.}\ \bibnamefont {Hong}}, \bibinfo {author} {\bibfnamefont
  {R.}~\bibnamefont {Li}}, \bibinfo {author} {\bibfnamefont {Q.}~\bibnamefont
  {Gao}}, \bibinfo {author} {\bibfnamefont {M.}~\bibnamefont {Chen}}, \bibinfo
  {author} {\bibfnamefont {W.}~\bibnamefont {Ren}}, \bibinfo {author}
  {\bibfnamefont {H.-M.}\ \bibnamefont {Cheng}}, \bibinfo {author}
  {\bibfnamefont {Y.}~\bibnamefont {Li}},\ and\ \bibinfo {author}
  {\bibfnamefont {X.-Q.}\ \bibnamefont {Chen}},\ }\href
  {https://doi.org/10.1038/s41467-021-22324-8} {\bibfield  {journal} {\bibinfo
  {journal} {Nature Communications}\ }\textbf {\bibinfo {volume} {12}},\
  \bibinfo {pages} {2361} (\bibinfo {year} {2021})}\BibitemShut {NoStop}%
\bibitem [{\citenamefont {Bafekry}\ \emph {et~al.}(2020)\citenamefont
  {Bafekry}, \citenamefont {Faraji}, \citenamefont {Hoat}, \citenamefont
  {Fadlallah}, \citenamefont {Shahrokhi}, \citenamefont {Shojaei},
  \citenamefont {Gogova},\ and\ \citenamefont
  {Ghergherehchi}}]{2020_MoSi2N4_ML_properties_arxiv}%
  \BibitemOpen
  \bibfield  {author} {\bibinfo {author} {\bibfnamefont {A.}~\bibnamefont
  {Bafekry}}, \bibinfo {author} {\bibfnamefont {M.}~\bibnamefont {Faraji}},
  \bibinfo {author} {\bibfnamefont {D.~M.}\ \bibnamefont {Hoat}}, \bibinfo
  {author} {\bibfnamefont {M.~M.}\ \bibnamefont {Fadlallah}}, \bibinfo {author}
  {\bibfnamefont {M.}~\bibnamefont {Shahrokhi}}, \bibinfo {author}
  {\bibfnamefont {F.}~\bibnamefont {Shojaei}}, \bibinfo {author} {\bibfnamefont
  {D.}~\bibnamefont {Gogova}},\ and\ \bibinfo {author} {\bibfnamefont
  {M.}~\bibnamefont {Ghergherehchi}},\ }\href {arxiv.org/abs/2009.04267}
  {\bibfield  {journal} {\bibinfo  {journal} {arXiv:2009.04267v1}\ ,\ \bibinfo
  {pages} {1}} (\bibinfo {year} {2020})}\BibitemShut {NoStop}%
\bibitem [{\citenamefont {Yu}\ \emph {et~al.}(2021)\citenamefont {Yu},
  \citenamefont {Zhou}, \citenamefont {Wan},\ and\ \citenamefont
  {Li}}]{2021_MoSi2N4_Kappa_NJP}%
  \BibitemOpen
  \bibfield  {author} {\bibinfo {author} {\bibfnamefont {J.}~\bibnamefont
  {Yu}}, \bibinfo {author} {\bibfnamefont {J.}~\bibnamefont {Zhou}}, \bibinfo
  {author} {\bibfnamefont {X.}~\bibnamefont {Wan}},\ and\ \bibinfo {author}
  {\bibfnamefont {Q.}~\bibnamefont {Li}},\ }\href
  {https://doi.org/10.1088/1367-2630/abe8f7} {\bibfield  {journal} {\bibinfo
  {journal} {New Journal of Physics}\ }\textbf {\bibinfo {volume} {23}},\
  \bibinfo {pages} {033005} (\bibinfo {year} {2021})}\BibitemShut {NoStop}%
\bibitem [{\citenamefont {Wu}\ \emph {et~al.}(2021)\citenamefont {Wu},
  \citenamefont {Cao}, \citenamefont {Ang},\ and\ \citenamefont
  {Ang}}]{2021_MoSi2N4_WSi2N4_MIT_APL}%
  \BibitemOpen
  \bibfield  {author} {\bibinfo {author} {\bibfnamefont {Q.}~\bibnamefont
  {Wu}}, \bibinfo {author} {\bibfnamefont {L.}~\bibnamefont {Cao}}, \bibinfo
  {author} {\bibfnamefont {Y.~S.}\ \bibnamefont {Ang}},\ and\ \bibinfo {author}
  {\bibfnamefont {L.~K.}\ \bibnamefont {Ang}},\ }\href
  {https://doi.org/10.1063/5.0044431} {\bibfield  {journal} {\bibinfo
  {journal} {Applied Physics Letters}\ }\textbf {\bibinfo {volume} {118}},\
  \bibinfo {pages} {113102} (\bibinfo {year} {2021})}\BibitemShut {NoStop}%
\bibitem [{\citenamefont {Zhong}\ \emph {et~al.}(2021)\citenamefont {Zhong},
  \citenamefont {Xiong}, \citenamefont {Lv}, \citenamefont {Yu},\ and\
  \citenamefont {Yuan}}]{2021_MA2Z4_MIT_Strain_PRB}%
  \BibitemOpen
  \bibfield  {author} {\bibinfo {author} {\bibfnamefont {H.}~\bibnamefont
  {Zhong}}, \bibinfo {author} {\bibfnamefont {W.}~\bibnamefont {Xiong}},
  \bibinfo {author} {\bibfnamefont {P.}~\bibnamefont {Lv}}, \bibinfo {author}
  {\bibfnamefont {J.}~\bibnamefont {Yu}},\ and\ \bibinfo {author}
  {\bibfnamefont {S.}~\bibnamefont {Yuan}},\ }\href
  {https://doi.org/10.1103/PhysRevB.103.085124} {\bibfield  {journal} {\bibinfo
   {journal} {Phys. Rev. B}\ }\textbf {\bibinfo {volume} {103}},\ \bibinfo
  {pages} {085124} (\bibinfo {year} {2021})}\BibitemShut {NoStop}%
\bibitem [{\citenamefont {Yao}\ \emph {et~al.}(2021)\citenamefont {Yao},
  \citenamefont {Zhang}, \citenamefont {Wang}, \citenamefont {Li},
  \citenamefont {Yu}, \citenamefont {Xu}, \citenamefont {Wang},\ and\
  \citenamefont {Wei}}]{2021_MoSe2Z4_opto_Nanomat}%
  \BibitemOpen
  \bibfield  {author} {\bibinfo {author} {\bibfnamefont {H.}~\bibnamefont
  {Yao}}, \bibinfo {author} {\bibfnamefont {C.}~\bibnamefont {Zhang}}, \bibinfo
  {author} {\bibfnamefont {Q.}~\bibnamefont {Wang}}, \bibinfo {author}
  {\bibfnamefont {J.}~\bibnamefont {Li}}, \bibinfo {author} {\bibfnamefont
  {Y.}~\bibnamefont {Yu}}, \bibinfo {author} {\bibfnamefont {F.}~\bibnamefont
  {Xu}}, \bibinfo {author} {\bibfnamefont {B.}~\bibnamefont {Wang}},\ and\
  \bibinfo {author} {\bibfnamefont {Y.}~\bibnamefont {Wei}},\ }\href
  {https://doi.org/10.3390/nano11030559} {\bibfield  {journal} {\bibinfo
  {journal} {Nanomaterials}\ }\textbf {\bibinfo {volume} {11}},\ \bibinfo
  {pages} {559} (\bibinfo {year} {2021})}\BibitemShut {NoStop}%
\bibitem [{\citenamefont {Kresse}\ and\ \citenamefont {Hafner}(1993)}]{VASP}%
  \BibitemOpen
  \bibfield  {author} {\bibinfo {author} {\bibfnamefont {G.}~\bibnamefont
  {Kresse}}\ and\ \bibinfo {author} {\bibfnamefont {J.}~\bibnamefont
  {Hafner}},\ }\href {https://link.aps.org/doi/10.1103/PhysRevB.47.558}
  {\bibfield  {journal} {\bibinfo  {journal} {Phys. Rev. B}\ }\textbf {\bibinfo
  {volume} {47}},\ \bibinfo {pages} {558} (\bibinfo {year} {1993})}\BibitemShut
  {NoStop}%
\bibitem [{\citenamefont {Blochl}(1994)}]{PAW}%
  \BibitemOpen
  \bibfield  {author} {\bibinfo {author} {\bibfnamefont {P.~E.}\ \bibnamefont
  {Blochl}},\ }\href {https://link.aps.org/doi/10.1103/PhysRevB.50.17953}
  {\bibfield  {journal} {\bibinfo  {journal} {Phys. Rev. B}\ }\textbf {\bibinfo
  {volume} {50}},\ \bibinfo {pages} {17953} (\bibinfo {year}
  {1994})}\BibitemShut {NoStop}%
\bibitem [{\citenamefont {Perdew}\ \emph {et~al.}(1996)\citenamefont {Perdew},
  \citenamefont {Burke},\ and\ \citenamefont {Ernzerhof}}]{PBE}%
  \BibitemOpen
  \bibfield  {author} {\bibinfo {author} {\bibfnamefont {P.}~\bibnamefont
  {Perdew}}, \bibinfo {author} {\bibfnamefont {K.}~\bibnamefont {Burke}},\ and\
  \bibinfo {author} {\bibfnamefont {M.}~\bibnamefont {Ernzerhof}},\ }\href
  {https://link.aps.org/doi/10.1103/PhysRevLett.77.3865} {\bibfield  {journal}
  {\bibinfo  {journal} {Phys. Rev. Lett.}\ }\textbf {\bibinfo {volume} {77}},\
  \bibinfo {pages} {3865} (\bibinfo {year} {1996})}\BibitemShut {NoStop}%
\bibitem [{\citenamefont {Dudarev}\ \emph {et~al.}(1998)\citenamefont
  {Dudarev}, \citenamefont {Botton}, \citenamefont {Savrasov}, \citenamefont
  {Humphreys},\ and\ \citenamefont {Sutton}}]{PBEU}%
  \BibitemOpen
  \bibfield  {author} {\bibinfo {author} {\bibfnamefont {S.~L.}\ \bibnamefont
  {Dudarev}}, \bibinfo {author} {\bibfnamefont {G.~A.}\ \bibnamefont {Botton}},
  \bibinfo {author} {\bibfnamefont {S.~Y.}\ \bibnamefont {Savrasov}}, \bibinfo
  {author} {\bibfnamefont {C.~J.}\ \bibnamefont {Humphreys}},\ and\ \bibinfo
  {author} {\bibfnamefont {A.~P.}\ \bibnamefont {Sutton}},\ }\href
  {https://doi.org/10.1103/PhysRevB.57.1505} {\bibfield  {journal} {\bibinfo
  {journal} {Phys. Rev. B}\ }\textbf {\bibinfo {volume} {57}},\ \bibinfo
  {pages} {1505} (\bibinfo {year} {1998})}\BibitemShut {NoStop}%
\bibitem [{\citenamefont {Heyd}\ \emph {et~al.}(2003)\citenamefont {Heyd},
  \citenamefont {Scuseria},\ and\ \citenamefont {Ernzerhof}}]{HSE03}%
  \BibitemOpen
  \bibfield  {author} {\bibinfo {author} {\bibfnamefont {J.}~\bibnamefont
  {Heyd}}, \bibinfo {author} {\bibfnamefont {G.~E.}\ \bibnamefont {Scuseria}},\
  and\ \bibinfo {author} {\bibfnamefont {M.}~\bibnamefont {Ernzerhof}},\ }\href
  {https://doi.org/10.1063/1.1564060} {\bibfield  {journal} {\bibinfo
  {journal} {The Journal of Chemical Physics}\ }\textbf {\bibinfo {volume}
  {118}},\ \bibinfo {pages} {8207} (\bibinfo {year} {2003})}\BibitemShut
  {NoStop}%
\bibitem [{\citenamefont {Krukau}\ \emph {et~al.}(2006)\citenamefont {Krukau},
  \citenamefont {Vydrov}, \citenamefont {Izmaylov},\ and\ \citenamefont
  {Scuseria}}]{HSE06}%
  \BibitemOpen
  \bibfield  {author} {\bibinfo {author} {\bibfnamefont {A.~V.}\ \bibnamefont
  {Krukau}}, \bibinfo {author} {\bibfnamefont {O.~A.}\ \bibnamefont {Vydrov}},
  \bibinfo {author} {\bibfnamefont {A.~F.}\ \bibnamefont {Izmaylov}},\ and\
  \bibinfo {author} {\bibfnamefont {G.~E.}\ \bibnamefont {Scuseria}},\ }\href
  {https://doi.org/10.1063/1.2404663} {\bibfield  {journal} {\bibinfo
  {journal} {The Journal of Chemical Physics}\ }\textbf {\bibinfo {volume}
  {125}},\ \bibinfo {pages} {224106} (\bibinfo {year} {2006})}\BibitemShut
  {NoStop}%
\bibitem [{\citenamefont {Kinaci}\ \emph {et~al.}(2015)\citenamefont {Kinaci},
  \citenamefont {Kado}, \citenamefont {Rosenmann}, \citenamefont {Ling},
  \citenamefont {Zhu}, \citenamefont {Banerjee},\ and\ \citenamefont
  {Chan}}]{HubbardUAPL}%
  \BibitemOpen
  \bibfield  {author} {\bibinfo {author} {\bibfnamefont {A.}~\bibnamefont
  {Kinaci}}, \bibinfo {author} {\bibfnamefont {M.}~\bibnamefont {Kado}},
  \bibinfo {author} {\bibfnamefont {D.}~\bibnamefont {Rosenmann}}, \bibinfo
  {author} {\bibfnamefont {C.}~\bibnamefont {Ling}}, \bibinfo {author}
  {\bibfnamefont {G.}~\bibnamefont {Zhu}}, \bibinfo {author} {\bibfnamefont
  {D.}~\bibnamefont {Banerjee}},\ and\ \bibinfo {author} {\bibfnamefont
  {M.~K.~Y.}\ \bibnamefont {Chan}},\ }\href {https://doi.org/10.1063/1.4938555}
  {\bibfield  {journal} {\bibinfo  {journal} {Applied Physics Letters}\
  }\textbf {\bibinfo {volume} {107}},\ \bibinfo {pages} {262108} (\bibinfo
  {year} {2015})}\BibitemShut {NoStop}%
\bibitem [{\citenamefont {Wang}\ \emph {et~al.}(2019)\citenamefont {Wang},
  \citenamefont {Puggioni},\ and\ \citenamefont {Rondinelli}}]{HubbardPRB}%
  \BibitemOpen
  \bibfield  {author} {\bibinfo {author} {\bibfnamefont {Y.}~\bibnamefont
  {Wang}}, \bibinfo {author} {\bibfnamefont {D.}~\bibnamefont {Puggioni}},\
  and\ \bibinfo {author} {\bibfnamefont {J.~M.}\ \bibnamefont {Rondinelli}},\
  }\href {https://doi.org/10.1103/PhysRevB.100.115149} {\bibfield  {journal}
  {\bibinfo  {journal} {Phys. Rev. B}\ }\textbf {\bibinfo {volume} {100}},\
  \bibinfo {pages} {115149} (\bibinfo {year} {2019})}\BibitemShut {NoStop}%
\bibitem [{\citenamefont {Zimmermann}\ \emph {et~al.}(2019)\citenamefont
  {Zimmermann}, \citenamefont {Bihlmayer}, \citenamefont {B\"ottcher},
  \citenamefont {Bouhassoune}, \citenamefont {Lounis}, \citenamefont {Sinova},
  \citenamefont {Heinze}, \citenamefont {Bl\"ugel},\ and\ \citenamefont
  {Dup\'e}}]{diffapproach}%
  \BibitemOpen
  \bibfield  {author} {\bibinfo {author} {\bibfnamefont {B.}~\bibnamefont
  {Zimmermann}}, \bibinfo {author} {\bibfnamefont {G.}~\bibnamefont
  {Bihlmayer}}, \bibinfo {author} {\bibfnamefont {M.}~\bibnamefont
  {B\"ottcher}}, \bibinfo {author} {\bibfnamefont {M.}~\bibnamefont
  {Bouhassoune}}, \bibinfo {author} {\bibfnamefont {S.}~\bibnamefont {Lounis}},
  \bibinfo {author} {\bibfnamefont {J.}~\bibnamefont {Sinova}}, \bibinfo
  {author} {\bibfnamefont {S.}~\bibnamefont {Heinze}}, \bibinfo {author}
  {\bibfnamefont {S.}~\bibnamefont {Bl\"ugel}},\ and\ \bibinfo {author}
  {\bibfnamefont {B.}~\bibnamefont {Dup\'e}},\ }\href
  {https://doi.org/10.1103/PhysRevB.99.214426} {\bibfield  {journal} {\bibinfo
  {journal} {Phys. Rev. B}\ }\textbf {\bibinfo {volume} {99}},\ \bibinfo
  {pages} {214426} (\bibinfo {year} {2019})}\BibitemShut {NoStop}%
\bibitem [{\citenamefont {Schweflinghaus}\ \emph {et~al.}(2016)\citenamefont
  {Schweflinghaus}, \citenamefont {Zimmermann}, \citenamefont {Heide},
  \citenamefont {Bihlmayer},\ and\ \citenamefont {Bl\"ugel}}]{diffapproachref}%
  \BibitemOpen
  \bibfield  {author} {\bibinfo {author} {\bibfnamefont {B.}~\bibnamefont
  {Schweflinghaus}}, \bibinfo {author} {\bibfnamefont {B.}~\bibnamefont
  {Zimmermann}}, \bibinfo {author} {\bibfnamefont {M.}~\bibnamefont {Heide}},
  \bibinfo {author} {\bibfnamefont {G.}~\bibnamefont {Bihlmayer}},\ and\
  \bibinfo {author} {\bibfnamefont {S.}~\bibnamefont {Bl\"ugel}},\ }\href
  {https://doi.org/10.1103/PhysRevB.94.024403} {\bibfield  {journal} {\bibinfo
  {journal} {Phys. Rev. B}\ }\textbf {\bibinfo {volume} {94}},\ \bibinfo
  {pages} {024403} (\bibinfo {year} {2016})}\BibitemShut {NoStop}%
\bibitem [{\citenamefont {Toyoda}\ \emph {et~al.}(2013)\citenamefont {Toyoda},
  \citenamefont {Yamauchi},\ and\ \citenamefont {Oguchi}}]{j1j2j3}%
  \BibitemOpen
  \bibfield  {author} {\bibinfo {author} {\bibfnamefont {M.}~\bibnamefont
  {Toyoda}}, \bibinfo {author} {\bibfnamefont {K.}~\bibnamefont {Yamauchi}},\
  and\ \bibinfo {author} {\bibfnamefont {T.}~\bibnamefont {Oguchi}},\ }\href
  {https://doi.org/10.1103/PhysRevB.87.224430} {\bibfield  {journal} {\bibinfo
  {journal} {Phys. Rev. B}\ }\textbf {\bibinfo {volume} {87}},\ \bibinfo
  {pages} {224430} (\bibinfo {year} {2013})}\BibitemShut {NoStop}%
\bibitem [{\citenamefont {Xiang}\ \emph {et~al.}(2011)\citenamefont {Xiang},
  \citenamefont {Kan}, \citenamefont {Wei}, \citenamefont {Whangbo},\ and\
  \citenamefont {Gong}}]{j12}%
  \BibitemOpen
  \bibfield  {author} {\bibinfo {author} {\bibfnamefont {H.~J.}\ \bibnamefont
  {Xiang}}, \bibinfo {author} {\bibfnamefont {E.~J.}\ \bibnamefont {Kan}},
  \bibinfo {author} {\bibfnamefont {S.-H.}\ \bibnamefont {Wei}}, \bibinfo
  {author} {\bibfnamefont {M.-H.}\ \bibnamefont {Whangbo}},\ and\ \bibinfo
  {author} {\bibfnamefont {X.~G.}\ \bibnamefont {Gong}},\ }\href
  {https://doi.org/10.1103/PhysRevB.84.224429} {\bibfield  {journal} {\bibinfo
  {journal} {Phys. Rev. B}\ }\textbf {\bibinfo {volume} {84}},\ \bibinfo
  {pages} {224429} (\bibinfo {year} {2011})}\BibitemShut {NoStop}%
\bibitem [{\citenamefont {Akanda}\ \emph {et~al.}(2020)\citenamefont {Akanda},
  \citenamefont {Park},\ and\ \citenamefont {Lake}}]{j13}%
  \BibitemOpen
  \bibfield  {author} {\bibinfo {author} {\bibfnamefont {M.~R.~K.}\
  \bibnamefont {Akanda}}, \bibinfo {author} {\bibfnamefont {I.~J.}\
  \bibnamefont {Park}},\ and\ \bibinfo {author} {\bibfnamefont {R.~K.}\
  \bibnamefont {Lake}},\ }\href {https://doi.org/10.1103/PhysRevB.102.224414}
  {\bibfield  {journal} {\bibinfo  {journal} {Phys. Rev. B}\ }\textbf {\bibinfo
  {volume} {102}},\ \bibinfo {pages} {224414} (\bibinfo {year}
  {2020})}\BibitemShut {NoStop}%
\bibitem [{vam()}]{vampire_code}%
  \BibitemOpen
  \href@noop {} {}\bibinfo {note} {{\mbox VAMPIRE software package version 5.0
  available from https://vampire.york.ac.uk}}\BibitemShut {NoStop}%
\bibitem [{\citenamefont {Evans}\ \emph {et~al.}(2014)\citenamefont {Evans},
  \citenamefont {Fan}, \citenamefont {Chureemart}, \citenamefont {Ostler},
  \citenamefont {Ellis},\ and\ \citenamefont {Chantrell}}]{vampire_JPCM14}%
  \BibitemOpen
  \bibfield  {author} {\bibinfo {author} {\bibfnamefont {R.~F.~L.}\
  \bibnamefont {Evans}}, \bibinfo {author} {\bibfnamefont {W.~J.}\ \bibnamefont
  {Fan}}, \bibinfo {author} {\bibfnamefont {P.}~\bibnamefont {Chureemart}},
  \bibinfo {author} {\bibfnamefont {T.~A.}\ \bibnamefont {Ostler}}, \bibinfo
  {author} {\bibfnamefont {M.~O.~A.}\ \bibnamefont {Ellis}},\ and\ \bibinfo
  {author} {\bibfnamefont {R.~W.}\ \bibnamefont {Chantrell}},\ }\href
  {https://doi.org/10.1088/0953-8984/26/10/103202} {\bibfield  {journal}
  {\bibinfo  {journal} {J. Phys.: Condens. Matter}\ }\textbf {\bibinfo {volume}
  {26}},\ \bibinfo {pages} {103202} (\bibinfo {year} {2014})}\BibitemShut
  {NoStop}%
\bibitem [{\citenamefont {Subhan}\ and\ \citenamefont
  {Hong}(2020)}]{VI3_Tc_JPCM20}%
  \BibitemOpen
  \bibfield  {author} {\bibinfo {author} {\bibfnamefont {F.}~\bibnamefont
  {Subhan}}\ and\ \bibinfo {author} {\bibfnamefont {J.}~\bibnamefont {Hong}},\
  }\href {https://doi.org/10.1088/1361-648x/ab7c14} {\bibfield  {journal}
  {\bibinfo  {journal} {Journal of Physics: Condensed Matter}\ }\textbf
  {\bibinfo {volume} {32}},\ \bibinfo {pages} {245803} (\bibinfo {year}
  {2020})}\BibitemShut {NoStop}%
\bibitem [{\citenamefont {Wang}\ \emph {et~al.}(2020)\citenamefont {Wang},
  \citenamefont {Qi},\ and\ \citenamefont {Qian}}]{CrSBrandCrSeBr}%
  \BibitemOpen
  \bibfield  {author} {\bibinfo {author} {\bibfnamefont {H.}~\bibnamefont
  {Wang}}, \bibinfo {author} {\bibfnamefont {J.}~\bibnamefont {Qi}},\ and\
  \bibinfo {author} {\bibfnamefont {X.}~\bibnamefont {Qian}},\ }\href
  {https://doi.org/10.1063/5.0014865} {\bibfield  {journal} {\bibinfo
  {journal} {Applied Physics Letters}\ }\textbf {\bibinfo {volume} {117}},\
  \bibinfo {pages} {083102} (\bibinfo {year} {2020})}\BibitemShut {NoStop}%
\bibitem [{\citenamefont {Takei}\ and\ \citenamefont
  {Tserkovnyak}(2014)}]{2014_SSF_YT_PRL}%
  \BibitemOpen
  \bibfield  {author} {\bibinfo {author} {\bibfnamefont {S.}~\bibnamefont
  {Takei}}\ and\ \bibinfo {author} {\bibfnamefont {Y.}~\bibnamefont
  {Tserkovnyak}},\ }\href {https://doi.org/10.1103/PhysRevLett.112.227201}
  {\bibfield  {journal} {\bibinfo  {journal} {Phys. Rev. Lett.}\ }\textbf
  {\bibinfo {volume} {112}},\ \bibinfo {pages} {227201} (\bibinfo {year}
  {2014})}\BibitemShut {NoStop}%
\bibitem [{\citenamefont {Fuh}\ \emph {et~al.}(2016)\citenamefont {Fuh},
  \citenamefont {Chang}, \citenamefont {Wang}, \citenamefont {Evans},
  \citenamefont {Chantrell},\ and\ \citenamefont {Jeng}}]{VS2andVSe2andVTe2}%
  \BibitemOpen
  \bibfield  {author} {\bibinfo {author} {\bibfnamefont {H.-R.}\ \bibnamefont
  {Fuh}}, \bibinfo {author} {\bibfnamefont {C.-R.}\ \bibnamefont {Chang}},
  \bibinfo {author} {\bibfnamefont {Y.-K.}\ \bibnamefont {Wang}}, \bibinfo
  {author} {\bibfnamefont {R.~F.~L.}\ \bibnamefont {Evans}}, \bibinfo {author}
  {\bibfnamefont {R.~W.}\ \bibnamefont {Chantrell}},\ and\ \bibinfo {author}
  {\bibfnamefont {H.-T.}\ \bibnamefont {Jeng}},\ }\href
  {https://doi.org/10.1038/srep32625} {\bibfield  {journal} {\bibinfo
  {journal} {Scientific Reports}\ }\textbf {\bibinfo {volume} {6}},\ \bibinfo
  {pages} {32625} (\bibinfo {year} {2016})}\BibitemShut {NoStop}%
\bibitem [{\citenamefont {Guan}\ and\ \citenamefont {Ni}(2020)}]{VSeTe}%
  \BibitemOpen
  \bibfield  {author} {\bibinfo {author} {\bibfnamefont {Z.}~\bibnamefont
  {Guan}}\ and\ \bibinfo {author} {\bibfnamefont {S.}~\bibnamefont {Ni}},\
  }\href {https://doi.org/10.1039/D0NR04837B} {\bibfield  {journal} {\bibinfo
  {journal} {Nanoscale}\ }\textbf {\bibinfo {volume} {12}},\ \bibinfo {pages}
  {22735} (\bibinfo {year} {2020})}\BibitemShut {NoStop}%
\bibitem [{\citenamefont {You}\ \emph {et~al.}(2020)\citenamefont {You},
  \citenamefont {Zhang}, \citenamefont {Dong}, \citenamefont {Gu},\ and\
  \citenamefont {Su}}]{MGeTe3}%
  \BibitemOpen
  \bibfield  {author} {\bibinfo {author} {\bibfnamefont {J.-Y.}\ \bibnamefont
  {You}}, \bibinfo {author} {\bibfnamefont {Z.}~\bibnamefont {Zhang}}, \bibinfo
  {author} {\bibfnamefont {X.-J.}\ \bibnamefont {Dong}}, \bibinfo {author}
  {\bibfnamefont {B.}~\bibnamefont {Gu}},\ and\ \bibinfo {author}
  {\bibfnamefont {G.}~\bibnamefont {Su}},\ }\href
  {https://doi.org/10.1103/PhysRevResearch.2.013002} {\bibfield  {journal}
  {\bibinfo  {journal} {Phys. Rev. Research}\ }\textbf {\bibinfo {volume}
  {2}},\ \bibinfo {pages} {013002} (\bibinfo {year} {2020})}\BibitemShut
  {NoStop}%
\bibitem [{\citenamefont {Zhang}\ \emph {et~al.}(2017)\citenamefont {Zhang},
  \citenamefont {Pang}, \citenamefont {Zhang}, \citenamefont {Gu},\ and\
  \citenamefont {Huang}}]{Co2S2}%
  \BibitemOpen
  \bibfield  {author} {\bibinfo {author} {\bibfnamefont {Y.}~\bibnamefont
  {Zhang}}, \bibinfo {author} {\bibfnamefont {J.}~\bibnamefont {Pang}},
  \bibinfo {author} {\bibfnamefont {M.}~\bibnamefont {Zhang}}, \bibinfo
  {author} {\bibfnamefont {X.}~\bibnamefont {Gu}},\ and\ \bibinfo {author}
  {\bibfnamefont {L.}~\bibnamefont {Huang}},\ }\href
  {https://doi.org/10.1038/s41598-017-16032-x} {\bibfield  {journal} {\bibinfo
  {journal} {Scientific Reports}\ }\textbf {\bibinfo {volume} {7}},\ \bibinfo
  {pages} {15993} (\bibinfo {year} {2017})}\BibitemShut {NoStop}%
\bibitem [{\citenamefont {Huang}\ \emph
  {et~al.}(2017{\natexlab{b}})\citenamefont {Huang}, \citenamefont {Zhou},
  \citenamefont {Wu}, \citenamefont {Deng}, \citenamefont {Jena},\ and\
  \citenamefont {Kan}}]{RuI3}%
  \BibitemOpen
  \bibfield  {author} {\bibinfo {author} {\bibfnamefont {C.}~\bibnamefont
  {Huang}}, \bibinfo {author} {\bibfnamefont {J.}~\bibnamefont {Zhou}},
  \bibinfo {author} {\bibfnamefont {H.}~\bibnamefont {Wu}}, \bibinfo {author}
  {\bibfnamefont {K.}~\bibnamefont {Deng}}, \bibinfo {author} {\bibfnamefont
  {P.}~\bibnamefont {Jena}},\ and\ \bibinfo {author} {\bibfnamefont
  {E.}~\bibnamefont {Kan}},\ }\href
  {https://doi.org/10.1103/PhysRevB.95.045113} {\bibfield  {journal} {\bibinfo
  {journal} {Phys. Rev. B}\ }\textbf {\bibinfo {volume} {95}},\ \bibinfo
  {pages} {045113} (\bibinfo {year} {2017}{\natexlab{b}})}\BibitemShut
  {NoStop}%
\bibitem [{\citenamefont {Wu}\ \emph {et~al.}(2015)\citenamefont {Wu},
  \citenamefont {Huang}, \citenamefont {Wu}, \citenamefont {Lee}, \citenamefont
  {Deng}, \citenamefont {Kan},\ and\ \citenamefont {Jena}}]{MoN2}%
  \BibitemOpen
  \bibfield  {author} {\bibinfo {author} {\bibfnamefont {F.}~\bibnamefont
  {Wu}}, \bibinfo {author} {\bibfnamefont {C.}~\bibnamefont {Huang}}, \bibinfo
  {author} {\bibfnamefont {H.}~\bibnamefont {Wu}}, \bibinfo {author}
  {\bibfnamefont {C.}~\bibnamefont {Lee}}, \bibinfo {author} {\bibfnamefont
  {K.}~\bibnamefont {Deng}}, \bibinfo {author} {\bibfnamefont {E.}~\bibnamefont
  {Kan}},\ and\ \bibinfo {author} {\bibfnamefont {P.}~\bibnamefont {Jena}},\
  }\href {https://doi.org/10.1021/acs.nanolett.5b03835} {\bibfield  {journal}
  {\bibinfo  {journal} {Nano Letters}\ }\textbf {\bibinfo {volume} {15}},\
  \bibinfo {pages} {8277} (\bibinfo {year} {2015})},\ \bibinfo {note} {pMID:
  26575002}\BibitemShut {NoStop}%
\bibitem [{\citenamefont {Xu}\ and\ \citenamefont {Zhu}(2018)}]{MnN}%
  \BibitemOpen
  \bibfield  {author} {\bibinfo {author} {\bibfnamefont {Z.}~\bibnamefont
  {Xu}}\ and\ \bibinfo {author} {\bibfnamefont {H.}~\bibnamefont {Zhu}},\
  }\href {https://doi.org/10.1021/acs.jpcc.8b02323} {\bibfield  {journal}
  {\bibinfo  {journal} {The Journal of Physical Chemistry C}\ }\textbf
  {\bibinfo {volume} {122}},\ \bibinfo {pages} {14918} (\bibinfo {year}
  {2018})}\BibitemShut {NoStop}%
\bibitem [{\citenamefont {Wang}\ \emph {et~al.}(2015)\citenamefont {Wang},
  \citenamefont {Ge}, \citenamefont {Sun}, \citenamefont {Zhang}, \citenamefont
  {Liu}, \citenamefont {Wen}, \citenamefont {Yu}, \citenamefont {Wang},
  \citenamefont {Zhang}, \citenamefont {Xu}, \citenamefont {Neuefeind},
  \citenamefont {Qin}, \citenamefont {Chen}, \citenamefont {Jin}, \citenamefont
  {Li}, \citenamefont {He},\ and\ \citenamefont
  {Zhao}}]{2015_3R-MoN2_Hi_P_Synthesis_JACS}%
  \BibitemOpen
  \bibfield  {author} {\bibinfo {author} {\bibfnamefont {S.}~\bibnamefont
  {Wang}}, \bibinfo {author} {\bibfnamefont {H.}~\bibnamefont {Ge}}, \bibinfo
  {author} {\bibfnamefont {S.}~\bibnamefont {Sun}}, \bibinfo {author}
  {\bibfnamefont {J.}~\bibnamefont {Zhang}}, \bibinfo {author} {\bibfnamefont
  {F.}~\bibnamefont {Liu}}, \bibinfo {author} {\bibfnamefont {X.}~\bibnamefont
  {Wen}}, \bibinfo {author} {\bibfnamefont {X.}~\bibnamefont {Yu}}, \bibinfo
  {author} {\bibfnamefont {L.}~\bibnamefont {Wang}}, \bibinfo {author}
  {\bibfnamefont {Y.}~\bibnamefont {Zhang}}, \bibinfo {author} {\bibfnamefont
  {H.}~\bibnamefont {Xu}}, \bibinfo {author} {\bibfnamefont {J.~C.}\
  \bibnamefont {Neuefeind}}, \bibinfo {author} {\bibfnamefont {Z.}~\bibnamefont
  {Qin}}, \bibinfo {author} {\bibfnamefont {C.}~\bibnamefont {Chen}}, \bibinfo
  {author} {\bibfnamefont {C.}~\bibnamefont {Jin}}, \bibinfo {author}
  {\bibfnamefont {Y.}~\bibnamefont {Li}}, \bibinfo {author} {\bibfnamefont
  {D.}~\bibnamefont {He}},\ and\ \bibinfo {author} {\bibfnamefont
  {Y.}~\bibnamefont {Zhao}},\ }\href {https://doi.org/10.1021/jacs.5b01446}
  {\bibfield  {journal} {\bibinfo  {journal} {Journal of the American Chemical
  Society}\ }\textbf {\bibinfo {volume} {137}},\ \bibinfo {pages} {4815}
  (\bibinfo {year} {2015})},\ \bibinfo {note} {pMID: 25799018}\BibitemShut
  {NoStop}%
\bibitem [{\citenamefont {Sorokin}\ \emph {et~al.}(2014)\citenamefont
  {Sorokin}, \citenamefont {Kvashnin}, \citenamefont {Zhu},\ and\ \citenamefont
  {Tománek}}]{2014_Spontaneous_Graphitization}%
  \BibitemOpen
  \bibfield  {author} {\bibinfo {author} {\bibfnamefont {P.~B.}\ \bibnamefont
  {Sorokin}}, \bibinfo {author} {\bibfnamefont {A.~G.}\ \bibnamefont
  {Kvashnin}}, \bibinfo {author} {\bibfnamefont {Z.}~\bibnamefont {Zhu}},\ and\
  \bibinfo {author} {\bibfnamefont {D.}~\bibnamefont {Tománek}},\ }\href
  {https://doi.org/10.1021/nl503673q} {\bibfield  {journal} {\bibinfo
  {journal} {Nano Letters}\ }\textbf {\bibinfo {volume} {14}},\ \bibinfo
  {pages} {7126} (\bibinfo {year} {2014})},\ \bibinfo {note} {pMID:
  25384500}\BibitemShut {NoStop}%
\bibitem [{\citenamefont {Webster}\ and\ \citenamefont {Yan}(2018)}]{MAECrX3}%
  \BibitemOpen
  \bibfield  {author} {\bibinfo {author} {\bibfnamefont {L.}~\bibnamefont
  {Webster}}\ and\ \bibinfo {author} {\bibfnamefont {J.-A.}\ \bibnamefont
  {Yan}},\ }\href {https://doi.org/10.1103/PhysRevB.98.144411} {\bibfield
  {journal} {\bibinfo  {journal} {Phys. Rev. B}\ }\textbf {\bibinfo {volume}
  {98}},\ \bibinfo {pages} {144411} (\bibinfo {year} {2018})}\BibitemShut
  {NoStop}%
\bibitem [{\citenamefont {Zhuang}\ \emph {et~al.}(2016)\citenamefont {Zhuang},
  \citenamefont {Kent},\ and\ \citenamefont {Hennig}}]{MAEFe3GeTe2}%
  \BibitemOpen
  \bibfield  {author} {\bibinfo {author} {\bibfnamefont {H.~L.}\ \bibnamefont
  {Zhuang}}, \bibinfo {author} {\bibfnamefont {P.~R.~C.}\ \bibnamefont
  {Kent}},\ and\ \bibinfo {author} {\bibfnamefont {R.~G.}\ \bibnamefont
  {Hennig}},\ }\href {https://doi.org/10.1103/PhysRevB.93.134407} {\bibfield
  {journal} {\bibinfo  {journal} {Phys. Rev. B}\ }\textbf {\bibinfo {volume}
  {93}},\ \bibinfo {pages} {134407} (\bibinfo {year} {2016})}\BibitemShut
  {NoStop}%
\bibitem [{\citenamefont {Torun}\ \emph {et~al.}(2015)\citenamefont {Torun},
  \citenamefont {Sahin}, \citenamefont {Bacaksiz}, \citenamefont {Senger},\
  and\ \citenamefont {Peeters}}]{MAEFeCl2}%
  \BibitemOpen
  \bibfield  {author} {\bibinfo {author} {\bibfnamefont {E.}~\bibnamefont
  {Torun}}, \bibinfo {author} {\bibfnamefont {H.}~\bibnamefont {Sahin}},
  \bibinfo {author} {\bibfnamefont {C.}~\bibnamefont {Bacaksiz}}, \bibinfo
  {author} {\bibfnamefont {R.~T.}\ \bibnamefont {Senger}},\ and\ \bibinfo
  {author} {\bibfnamefont {F.~M.}\ \bibnamefont {Peeters}},\ }\href
  {https://doi.org/10.1103/PhysRevB.92.104407} {\bibfield  {journal} {\bibinfo
  {journal} {Phys. Rev. B}\ }\textbf {\bibinfo {volume} {92}},\ \bibinfo
  {pages} {104407} (\bibinfo {year} {2015})}\BibitemShut {NoStop}%
\bibitem [{\citenamefont {Zheng}\ \emph {et~al.}(2019)\citenamefont {Zheng},
  \citenamefont {Huang}, \citenamefont {Yu}, \citenamefont {Xu}, \citenamefont
  {Zhang}, \citenamefont {Xu}, \citenamefont {Liu}, \citenamefont {Kan},
  \citenamefont {Wang},\ and\ \citenamefont {Yang}}]{MAEFe3P}%
  \BibitemOpen
  \bibfield  {author} {\bibinfo {author} {\bibfnamefont {S.}~\bibnamefont
  {Zheng}}, \bibinfo {author} {\bibfnamefont {C.}~\bibnamefont {Huang}},
  \bibinfo {author} {\bibfnamefont {T.}~\bibnamefont {Yu}}, \bibinfo {author}
  {\bibfnamefont {M.}~\bibnamefont {Xu}}, \bibinfo {author} {\bibfnamefont
  {S.}~\bibnamefont {Zhang}}, \bibinfo {author} {\bibfnamefont
  {H.}~\bibnamefont {Xu}}, \bibinfo {author} {\bibfnamefont {Y.}~\bibnamefont
  {Liu}}, \bibinfo {author} {\bibfnamefont {E.}~\bibnamefont {Kan}}, \bibinfo
  {author} {\bibfnamefont {Y.}~\bibnamefont {Wang}},\ and\ \bibinfo {author}
  {\bibfnamefont {G.}~\bibnamefont {Yang}},\ }\href
  {https://doi.org/10.1021/acs.jpclett.9b00970} {\bibfield  {journal} {\bibinfo
   {journal} {The Journal of Physical Chemistry Letters}\ }\textbf {\bibinfo
  {volume} {10}},\ \bibinfo {pages} {2733} (\bibinfo {year}
  {2019})}\BibitemShut {NoStop}%
\bibitem [{\citenamefont {Han}\ \emph {et~al.}(2020{\natexlab{b}})\citenamefont
  {Han}, \citenamefont {Zheng}, \citenamefont {Wang},\ and\ \citenamefont
  {Yan}}]{MAENiI2}%
  \BibitemOpen
  \bibfield  {author} {\bibinfo {author} {\bibfnamefont {H.}~\bibnamefont
  {Han}}, \bibinfo {author} {\bibfnamefont {H.}~\bibnamefont {Zheng}}, \bibinfo
  {author} {\bibfnamefont {Q.}~\bibnamefont {Wang}},\ and\ \bibinfo {author}
  {\bibfnamefont {Y.}~\bibnamefont {Yan}},\ }\href
  {https://doi.org/10.1039/D0CP03803B} {\bibfield  {journal} {\bibinfo
  {journal} {Phys. Chem. Chem. Phys.}\ }\textbf {\bibinfo {volume} {22}},\
  \bibinfo {pages} {26917} (\bibinfo {year} {2020}{\natexlab{b}})}\BibitemShut
  {NoStop}%
\bibitem [{\citenamefont {Park}\ \emph {et~al.}(2020)\citenamefont {Park},
  \citenamefont {Kwon},\ and\ \citenamefont {Lake}}]{IJPark_CrSb_PRB20}%
  \BibitemOpen
  \bibfield  {author} {\bibinfo {author} {\bibfnamefont {I.~J.}\ \bibnamefont
  {Park}}, \bibinfo {author} {\bibfnamefont {S.}~\bibnamefont {Kwon}},\ and\
  \bibinfo {author} {\bibfnamefont {R.~K.}\ \bibnamefont {Lake}},\ }\href
  {https://doi.org/10.1103/PhysRevB.102.224426} {\bibfield  {journal} {\bibinfo
   {journal} {Phys. Rev. B}\ }\textbf {\bibinfo {volume} {102}},\ \bibinfo
  {pages} {224426} (\bibinfo {year} {2020})}\BibitemShut {NoStop}%
\bibitem [{\citenamefont {Peng}\ \emph {et~al.}(2014)\citenamefont {Peng},
  \citenamefont {Wei},\ and\ \citenamefont {Copple}}]{HighStrain}%
  \BibitemOpen
  \bibfield  {author} {\bibinfo {author} {\bibfnamefont {X.}~\bibnamefont
  {Peng}}, \bibinfo {author} {\bibfnamefont {Q.}~\bibnamefont {Wei}},\ and\
  \bibinfo {author} {\bibfnamefont {A.}~\bibnamefont {Copple}},\ }\href
  {https://doi.org/10.1103/PhysRevB.90.085402} {\bibfield  {journal} {\bibinfo
  {journal} {Phys. Rev. B}\ }\textbf {\bibinfo {volume} {90}},\ \bibinfo
  {pages} {085402} (\bibinfo {year} {2014})}\BibitemShut {NoStop}%
\bibitem [{\citenamefont {Towns}\ \emph {et~al.}(2014)\citenamefont {Towns},
  \citenamefont {Cockerill}, \citenamefont {Dahan}, \citenamefont {Foster},
  \citenamefont {Gaither}, \citenamefont {Grimshaw}, \citenamefont {Hazlewood},
  \citenamefont {Lathrop}, \citenamefont {Lifka}, \citenamefont {Peterson}
  \emph {et~al.}}]{towns2014xsede}%
  \BibitemOpen
  \bibfield  {author} {\bibinfo {author} {\bibfnamefont {J.}~\bibnamefont
  {Towns}}, \bibinfo {author} {\bibfnamefont {T.}~\bibnamefont {Cockerill}},
  \bibinfo {author} {\bibfnamefont {M.}~\bibnamefont {Dahan}}, \bibinfo
  {author} {\bibfnamefont {I.}~\bibnamefont {Foster}}, \bibinfo {author}
  {\bibfnamefont {K.}~\bibnamefont {Gaither}}, \bibinfo {author} {\bibfnamefont
  {A.}~\bibnamefont {Grimshaw}}, \bibinfo {author} {\bibfnamefont
  {V.}~\bibnamefont {Hazlewood}}, \bibinfo {author} {\bibfnamefont
  {S.}~\bibnamefont {Lathrop}}, \bibinfo {author} {\bibfnamefont
  {D.}~\bibnamefont {Lifka}}, \bibinfo {author} {\bibfnamefont {G.~D.}\
  \bibnamefont {Peterson}}, \emph {et~al.},\ }\href
  {https://doi.org/10.1109/MCSE.2014.80} {\bibfield  {journal} {\bibinfo
  {journal} {Computing in Science \& Engineering}\ }\textbf {\bibinfo {volume}
  {16}},\ \bibinfo {pages} {62} (\bibinfo {year} {2014})}\BibitemShut {NoStop}%
\end{thebibliography}%

\end{document}

\title{Magnetic Properties of NbSi\textsubscript{2}N\textsubscript{4}, VSi\textsubscript{2}N\textsubscript{4}, and VSi\textsubscript{2}P\textsubscript{4} Monolayers: Supplementary Information}

\author{Md. Rakibul Karim Akanda}
\affiliation{Laboratory for Terahertz and Terascale Electronics (LATTE), Department of Electrical and Computer Engineering, University of California, Riverside, CA 92521, USA}

\author{Roger K. Lake}
\affiliation{Laboratory for Terahertz and Terascale Electronics (LATTE), Department of Electrical and Computer Engineering, University of California, Riverside, CA 92521, USA}
%\affiliation{Center of Spins and Heat in Nanoscale Electronic systems, University of California, Riverside, CA 92521, USA}

\date{\today}

\pacs{}% insert suggested PACS numbers in braces on next line

\maketitle %\maketitle must follow title, authors, abstract and \pacs

\newpage

\begin{figure}[H]
\centering
\includegraphics[width=.32\linewidth]{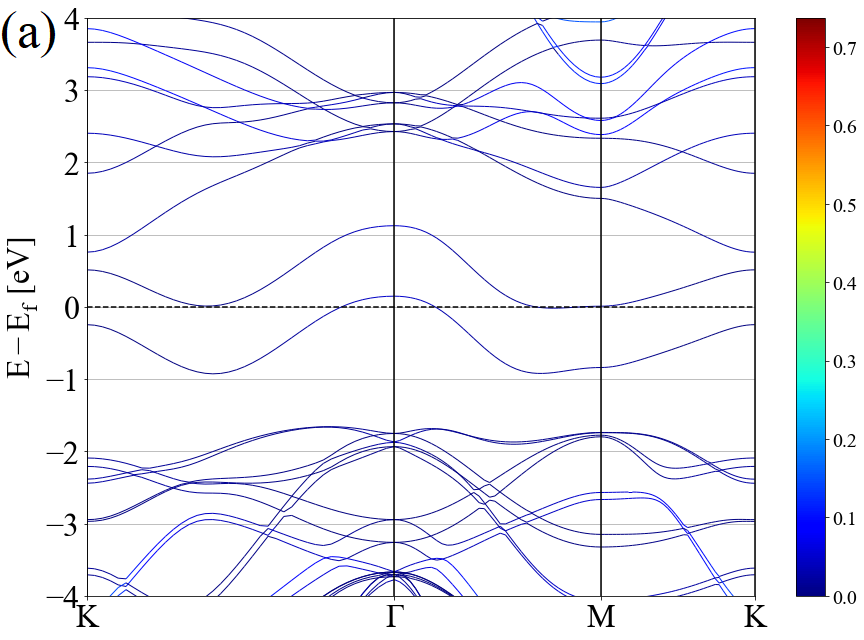}
\includegraphics[width=.32\linewidth]{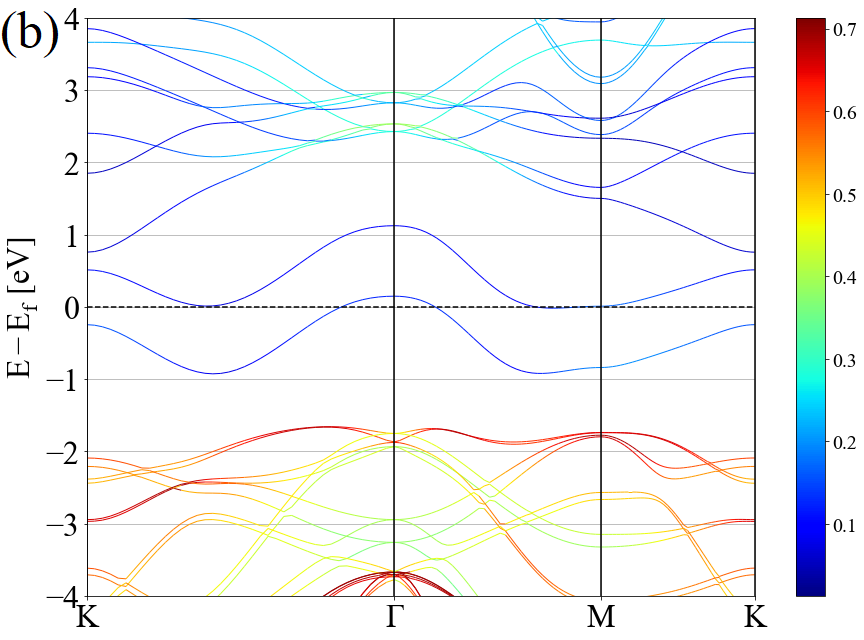}
\includegraphics[width=.32\linewidth]{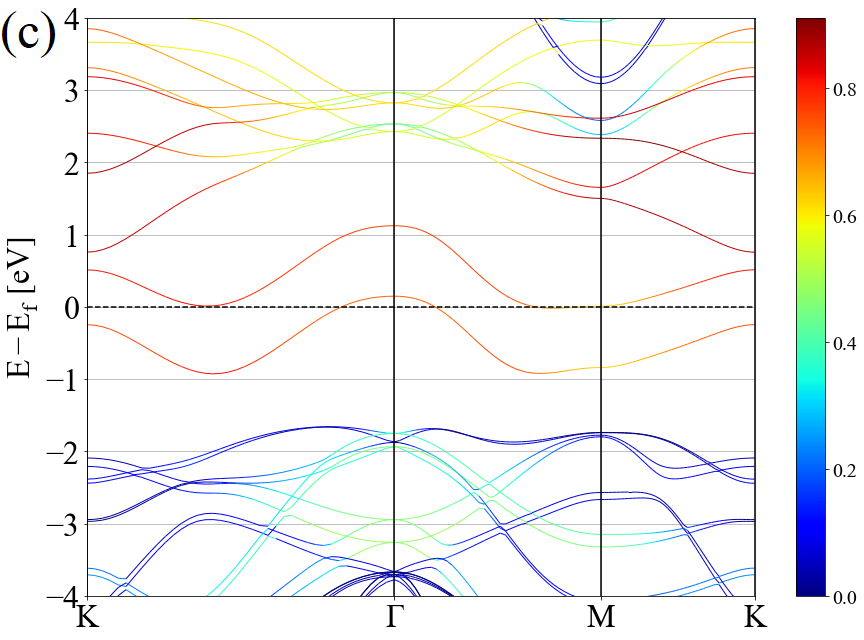}
\caption{\label{fig:Ek66}
Orbital resolved PBE band structure of VSi\textsubscript{2}N\textsubscript{4} at equilibrium:
(a) s-orbital contribution,
(b) p-orbital contribution, and
(c) d-orbital contribution.
The weight is given by the color bars at right.
The two isolated, narrow bands near the Fermi level are
d-orbital bands centered on the transition metal atoms vanadium (V).
The higher valence bands are primarily p-orbital bands which come from the N atoms.
The lower conduction bands are primarily d-orbital bands.
}
\end{figure}

\begin{figure}[H]
\centering
\includegraphics[width=.96\linewidth]{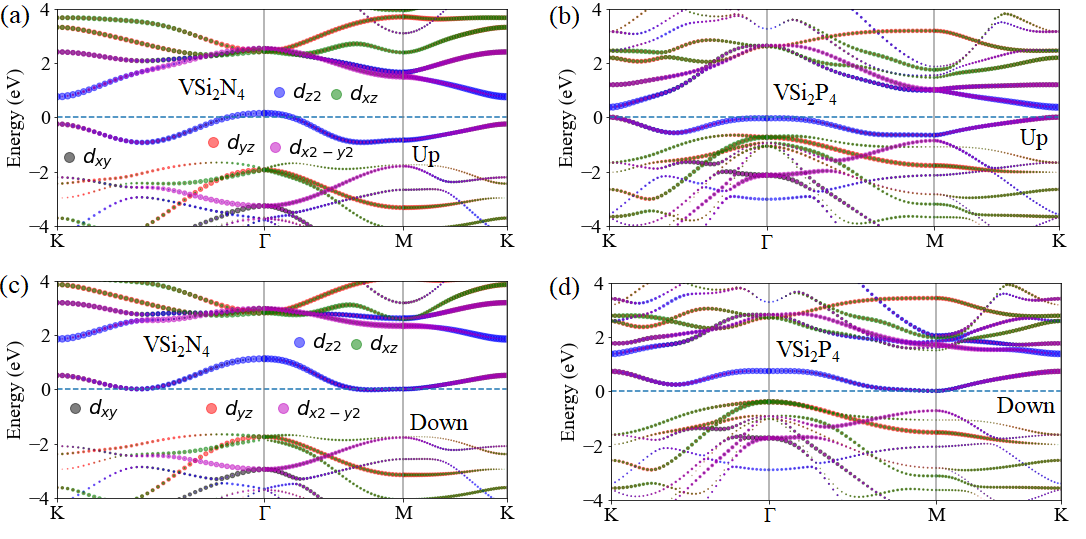}
\caption{\label{fig:dVSi2N4}
d-orbital resolved PBE band structure of VSi\textsubscript{2}N\textsubscript{4}: (a) spin up bands,
(c) spin down bands.
d-orbital resolved band structure of VSi\textsubscript{2}P\textsubscript{4}: (b) spin up bands, (d) spin down bands.
}
\end{figure}

\begin{figure}[H]
\centering
\includegraphics[width=.32\linewidth]{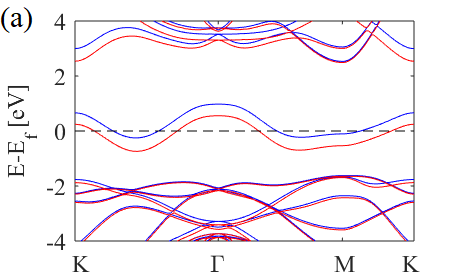}
\includegraphics[width=.32\linewidth]{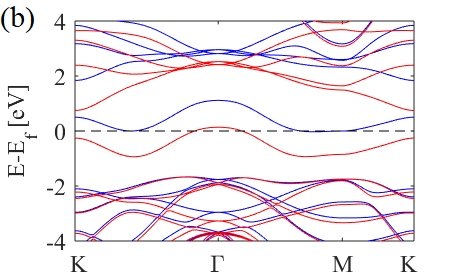}
\includegraphics[width=.32\linewidth]{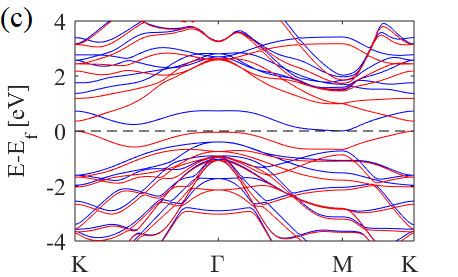}
\includegraphics[width=.32\linewidth]{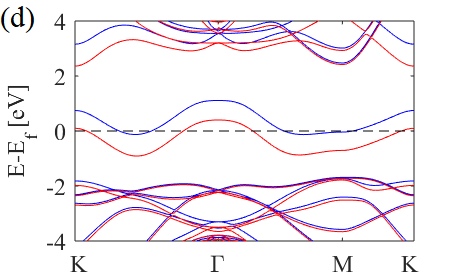}
\includegraphics[width=.32\linewidth]{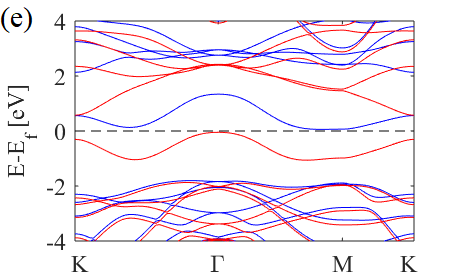}
\includegraphics[width=.32\linewidth]{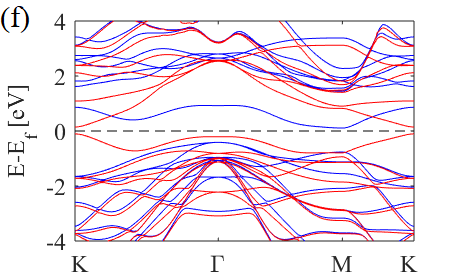}
\includegraphics[width=.32\linewidth]{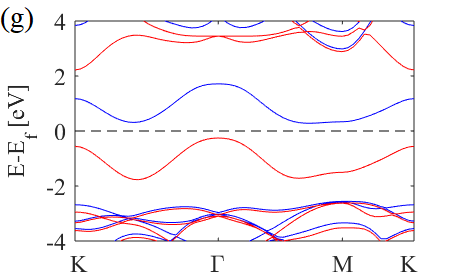}
\includegraphics[width=.32\linewidth]{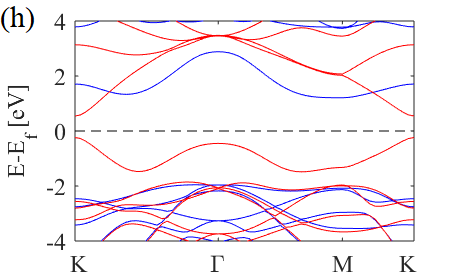}
\includegraphics[width=.32\linewidth]{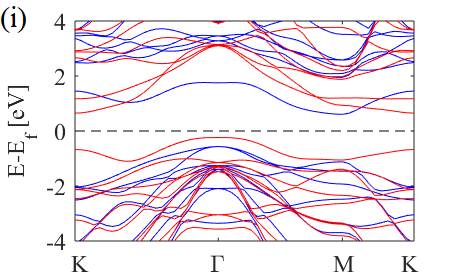}
\caption{\label{fig:EkU}
Spin resolved PBE band structure: (a) NbSi\textsubscript{2}N\textsubscript{4},
(b) VSi\textsubscript{2}N\textsubscript{4}, and
(c) VSi\textsubscript{2}P\textsubscript{4}.
Spin resolved PBE+U band structure: (d) NbSi\textsubscript{2}N\textsubscript{4},
(e) VSi\textsubscript{2}N\textsubscript{4}, and
(f) VSi\textsubscript{2}P\textsubscript{4}.
Spin resolved HSE06 band structure: (g) NbSi\textsubscript{2}N\textsubscript{4},
(h) VSi\textsubscript{2}N\textsubscript{4}, and
(i) VSi\textsubscript{2}P\textsubscript{4}.
}
\end{figure}

\begin{figure}[H]
\centering
\includegraphics[width=.32\linewidth]{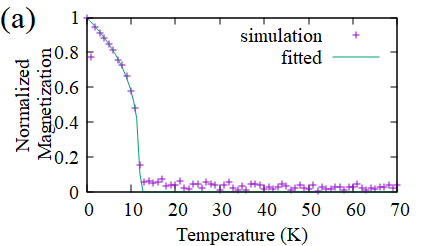}
\includegraphics[width=.32\linewidth]{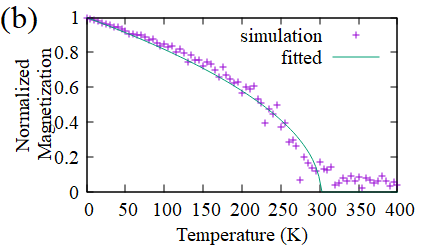}
\includegraphics[width=.32\linewidth]{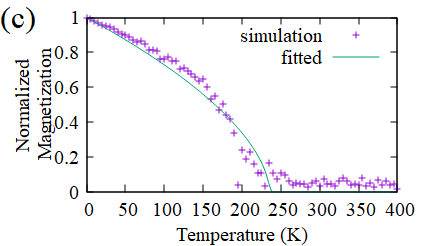}
\caption{\label{fig:UCurie}
MC calculations of the normalized magnetization as a function of temperature with
exchange constants extracted from PBE DFT for (a) NbSi\textsubscript{2}N\textsubscript{4},
(b) VSi\textsubscript{2}N\textsubscript{4}, and
(c) VSi\textsubscript{2}P\textsubscript{4}.
Curie temperatures are calculated from Monte Carlo simulation using the VAMPIRE software.}
\end{figure}

Fig.~\ref{fig:Ek66} shows the orbital resolved band structure of
VSi\textsubscript{2}N\textsubscript{4}.
Fig.~\ref{fig:dVSi2N4} shows the d-orbital resolved band structures of
VSi\textsubscript{2}N\textsubscript{4} and VSi\textsubscript{2}P\textsubscript{4}.
Spin resolved PBE, PBE+U, and HSE06 band structures are shown in Fig.~\ref{fig:EkU}.
Monte Carlo calculations using exchange constants extracted from PBE DFT are shown in Fig.~\ref{fig:UCurie}.

